\documentclass[utf8]{article} 

\usepackage{url,hyperref,lineno,microtype,subcaption}
\usepackage[onehalfspacing]{setspace}
\usepackage[margin=1in]{geometry}
\usepackage{natbib}
\usepackage{amsmath}
\usepackage{graphicx}

\usepackage{authblk}

\begin{document}
\onecolumn

\title{Granular packing simulation protocols: tap, press and relax} 
\author[1]{A. P. Santos\thanks{andrew.p.santos@nasa.gov}}
\author[2]{Ishan Srivastava}
\author[3]{Leonardo E. Silbert}
\author[4]{Jeremy B. Lechman}
\author[4]{Gary S. Grest}
\affil[1]{AMA Inc., Thermal Protection Materials Branch, NASA Ames Research Center, Moffett Field, CA 94035, USA.}
\affil[2]{Center for Computational Sciences and Engineering, Lawrence Berkeley National Laboratory, Berkeley, California 94720, USA.}
\affil[3]{School of Math, Science and Engineering, Central New Mexico Community College, Albuquerque, NM 87106, USA.}
\affil[4]{Sandia National Laboratories, Albuquerque, NM 87185, USA.}

\maketitle
\begin{abstract}
Granular matter takes many paths to pack. 
Gentle compression, compaction or repetitive tapping can happen in natural and industrial processes. 
The path influences the packing microstructure, and thus macroscale properties, particularly for frictional grains. 
We perform discrete element modeling simulations to construct packings of frictional spheres implementing a range of stress-controlled protocols with 3D periodic boundary conditions. 
A volume-controlled over-compression method is compared to four stress-controlled methods, including over-compression and release, gentle under-compression and cyclical compression and release. 
The packing volume fraction of each method depends on the pressure, initial kinetic energy and protocol parameters. 
A non-monotonic pressure dependence in the volume fraction, but not the coordination number occurs when dilute particles initialized with a non-zero kinetic energy are compressed, but can be reduced with the inclusion of drag.
The fraction of frictional contacts correlates with the volume fraction minimum.
Packings were cyclically compressed 1000 times. Response to compression depends on pressure; low pressure packings have a constant volume fraction regime, while high pressure
packings continue to get dense with number of cycles. 
The capability of stress-controlled, bulk-like particle simulations to capture different protocols is showcased, and the ability to pack at low pressures demonstrates unexpected behavior.
\end{abstract}


\renewcommand*\rmdefault{bch}\normalfont\upshape
\rmfamily
\section*{}
\vspace{-1cm}

\section{Introduction}\label{sec:intro}
Packings of granular materials are relevant to many industrial processes and natural phenomena.
Prediction and control of particle packing in industrial processes for particulate materials significantly impacts assurance, such as additively-manufactured part strength~\citep{Snow2019,Wischeropp2019}.
Packings formed naturally also depend on the packing process. For example, cut or fallen trees aggregation can improve stream restoration~\citep{Gerhard2000}, or damage bridges~\citep{Melville1988}.
Understanding the complex response of these far-from-equilibrium systems is critical to developing more efficient and effective means of controlling them.
Modeling is a powerful tool for deducing how controls affect the response of granular material processes. 
Access to particle-scale information, such as particle-particle forces, makes simulations well-equipped to study the effect of control in many phenomena. Simulations have shown that frictionless sphere packings approach the maximally random jammed state volume fraction~\citep{Torquato2000} and the coordination number set by isostaticity~\citep{OHern2003} for many different packing protocols. 
However the jamming point depends on material-specific contact mechanics and path to jamming~\citep{Luding2016}.
Frictionless particle packings can lead to various packing fractions by protocol changes in isotropic compression~\citep{Chaudhuri2010} or by applying shear strains~\citep{Bertrand2016}. Real granular particles have friction, and can form looser packings than frictionless sphere packings~\citep{Onoda1990,Silbert2010,santosGranularPackingsSliding2020}. Frictional particles can also access a range of packing fraction depending on the protocol. ~\citet{Song2008a} attribute the range of packing fractions as sampling an ensemble of jammed states in a statistical mechanical definition of jamming. Friction changes packing behavior beyond the volume fraction. For example, in 3 dimensions, the coordination number decreases gradually from the frictionless value of $Z=6$ to the frictional isostatic number $Z$=4 as the friction coefficient increases~\citep{Silbert2002,Shundyak2007,Somfai2007,Song2008a,Silbert2010}. The coordination number $Z$ and volume fraction $\phi$ of stable packings of particles with a specific friction coefficient can also depend on the path to packing~\citep{Silbert2002,Somfai2007,Bi2011}. 

Packings of granular particles can be formed many different ways. ~\citet{Farrell2010} formed low-density packings by settling granular particles in near-density-matched solvent. 
Applying drag in simulations, either to the simulation cell or the particles, forms low-density packings as do the near-density-matched solvent experiments~\citep{Delaney2011,hoyUnifiedAnalyticExpressions2020}. 
~\citet{Bililign2019} observed protocol dependence in experiments of two-dimensional packings under various protocols, for example uni- and bi-axial compression.
A common method to create dense particle packings is by isotropic compression. Volume-controlled compression can be achieved by randomly distributing point particles in a simulation cell and increasing the diameter~\citep{Lubachevsky1990,Shundyak2007}, or by decreasing the simulation cell density of an over-compressed system, while minimizing the conformational energy~\citep{OHern2002,Charbonneau2012}.
Flowing particles coming to a stop is another way for them to pack, for example from flow down an incline~\citep{Silbert2002} or by applying shear~\citep{Bi2011,Srivastava2019} or more complex flow geometries~\citep{clemmerShearNotAlways2021}. 
Path changes are also common methods, such as tapping or cyclical shear. 
These repetitive processes generally lead to denser packings~\citep{Kohlrausch1854,Williams1970,Knight1995,Philippe2002,Richard2005,Rosato2010,Kumar2016}.
The diversity of these research protocols is small compared to empirically developed protocols for industrial processes.

Simulation packing methods often control the volume, not the stress. Achieving zero-stress stable packings, for example, is difficult for such methods.
Previous jamming studies of particles with sliding friction as a function of pressure demonstrated that the packing fraction and the coordination number decrease monotonically with decreasing pressure~\citep{Shundyak2007,Silbert2010}.  
In this article, a constant pressure in the x-, y- and z-directions allows the simulation cell to adjust the edge length, and constant zero shear stresses allow the simulation cell to adopt triclinic configurations. The final packings repeatably and rigorously satisfy those stress conditions. ~\citet{Dagois-Bohy2012,Smith2014} showed that packings formed by controlling the pressure are more stable to shear deformation than volume-controlled methods. Furthermore, very low pressures are accessible to this protocol without extrapolation, unlike previous protocols~\citep{Silbert2010}. The method simulates a representative subset of particles, far from boundaries, in a real granular packing.
Similar protocols have been applied to 2D frictionless~\citep{Dagois-Bohy2012}, 3D frictionless~\citep{Smith2014}, 2D frictional~\citep{Shundyak2007,Somfai2007} and 3D frictional~\citep{santosGranularPackingsSliding2020} granular particles.

The equations of motion that describe this methodology are in Sec.~\ref{sec:boxeom}. The variety of packing methods available with pressure control are explored in Sec.~\ref{sec:packingmethod} and tested in Sec.~\ref{sec:protocols}. The low pressures that are accessible with this protocol highlight anomalous dense packings with low average coordination numbers. Sec.~\ref{sec:explanation} includes analysis of the resulting packings.

\section{Methodology}\label{sec:methods} 
\subsection{Constant stress simulations and particle model}\label{sec:boxeom}
Granular particles are modeled as spheres. Particles only interact when in contact, through a Hookean spring-dashpot-slider interaction potential, and they all have diameter $d$ and mass $m$. 
The particle spring and damping parameters are set equal to each other $k_n = k_s = 1$ and $\gamma_n = \gamma_s = 0.5 \tau^{-1}$ where $\tau = \sqrt{m/k_n}$ is the unit of time. 
The unit of pressure is $k_n/d$ and applies to all stresses; the unit of force is $k_nd$.
The assumption of linear elastic behavior for inter-particle contacts is reasonably accurate as a model for sufficiently stiff particles. 

Discrete element method (DEM) simulations, with the contact model described in Sec. \ref{sec:boxeom}, were performed using LAMMPS~\citep{thompsonLAMMPSFlexibleSimulation2022}.
The inter-particle forces $\mathbf{F}_i$ and torques $\boldsymbol{\tau}_i$ are used to integrate the equations of motion and update particle positions and orientations. To simulate granular particles under constant stress, the equations of motion include the degrees of freedom for a deforming simulation cell. The granular particles are placed within a periodic three-dimensional simulation cell that maintains an applied stress tensor by  making triclinic cell deformations. 
In particular, the Shinoda-Shiga-Mikami~\citep{Shinoda2004} formulation of a barostat was used to integrate the positions and momenta of the particles and to maintain an applied pressure tensor by varying the simulation cell. This formulation combines the hydrostatic equations of Martyna \textit{et al.} with the strain energy proposed by Parrinello and Rahman~\citep{Parrinello1981,Martyna1994},
\begin{subequations}
\begin{align}
\mathbf{\dot{r}}_i =& \frac{\mathbf{{p}}_i }{m} +\frac{\mathbf{{p}}_{\text{cell}} }{\omega_{\text{cell}}}\mathbf{{r}}_i \label{eqn:nph}\\
\mathbf{\dot{p}}_i =& \mathbf{F}_i - \frac{\mathbf{{p}}_{\text{cell}} -\frac{1}{N_f}\text{Tr}[\mathbf{{p}}_{\text{cell}}] }{\omega_{\text{cell}}}\mathbf{{p}}_i \\
\mathbf{\dot{h}} =& \frac{\mathbf{{p}}_{\text{cell}} }{\omega_{\text{cell}}}\mathbf{h} \\
\frac{\mathbf{\dot{p}}_{\text{cell}}}{k_{\text{drag}}} =& V(\mathbf{P_{\text{int}}-P_{{a}}}) - \mathbf{{h}}\Sigma\mathbf{h}^T + \frac{1}{N_f}\sum_{i=1}^{N}\frac{\mathbf{{p}}_i^2 }{m}\mathbf{I} \\
\omega_{\text{cell}} =& N\epsilon P_{\text{damp}}^2 \label{eqn:tdamp}
\end{align}
\end{subequations}
where $\mathbf{r}_i$ and $\mathbf{p}_i$ are the position and momentum vectors of the $i^{\text{th}}$ particle. 
A ``cell'' subscript refers to the simulation cell mass and momentum. The simulation cell ``momentum'' is modularly invariant, and has $\frac{md^2}{\tau}$ units.
$\mathbf{I}$ is the identity matrix, $V$ is the simulation cell volume, $\mathbf{P_{{a}}}$ is the applied pressure tensor and $\mathbf{P_{\text{int}}}$ is the internal pressure tensor. 	
The simulation cell ``mass'' $\omega_{\text{cell}}$ has units $m d^2$. Fluctuations in $\mathbf{P}_{\text{int}}$ as the system approaches $P_{a}$ are dampened by $P_{\text{damp}}$ which has units of $\tau$. The energy scale $\epsilon = 1 k_n$. As an athermal system, DEM simulations using this barostat ignore contributions typical to molecular dynamics simulations, such as thermostat chains\footnote[1]{To exclude thermostat chain and options in LAMMPS~\citep{thompsonLAMMPSFlexibleSimulation2022}, add \texttt{pchain 0 ptemp 1} to the \texttt{fix nph/sphere} barostat options.}~\citep{Shinoda2004}.

The triclinic deformations are captured by the simulation cell matrix $\mathbf{{h}}$. 
The $\mathbf{{h}}\Sigma\mathbf{{h^T}}$ term comes from the Parrinello-Rahman formulation~\citep{Parrinello1981} and represents the external applied stress, defined by reference matrix $\mathbf{{h}}_0$, where $\Sigma = \mathbf{h}_0^{-1}\left(\mathbf{P_{\text{int}}-P_{{a}}}\right)\mathbf{h}_0^{T-1}$. 
The internal pressure tensor $\mathbf{P_{\text{int}}}$ components
\begin{eqnarray}
P_{\text{int}}^{\alpha,\beta} = \frac{1}{V}\left[\sum_{i=1}^{N}\frac{\mathbf{{p}}_i^{\alpha}\mathbf{{p}}_i^{\beta}}{m} + \mathbf{F}_i^{\alpha} \mathbf{r}_i^{\beta} \right] 
\end{eqnarray}.
At jamming $\mathbf{P_{\text{int}}}=\mathbf{P_{a}}$ within numerical precision.
A computational, unitless drag factor $k_{\text{drag}}$ scales the simulation cell acceleration:
\begin{eqnarray}
k_{\text{drag}} =& 1 - \frac{\Delta t f_{\text{drag}}}{P_{\text{damp}}} 
\end{eqnarray}
where $\Delta t$ is the time step and $f_{\text{drag}}$ is a nonnegative, unitless input parameter\footnote[2]{Add \texttt{drag} $f_{\text{drag}}$ to the \texttt{fix nph/sphere} options to apply drag on the barostat in LAMMPS~\citep{thompsonLAMMPSFlexibleSimulation2022}.}. The simulation cell drag factor can mimic experimental packing protocols, or ensure stability flow simulations. 

\subsection{Packing methodology}\label{sec:packingmethod}
\begin{figure}[htbp]
    \begin{center}
    \includegraphics[width=9cm]{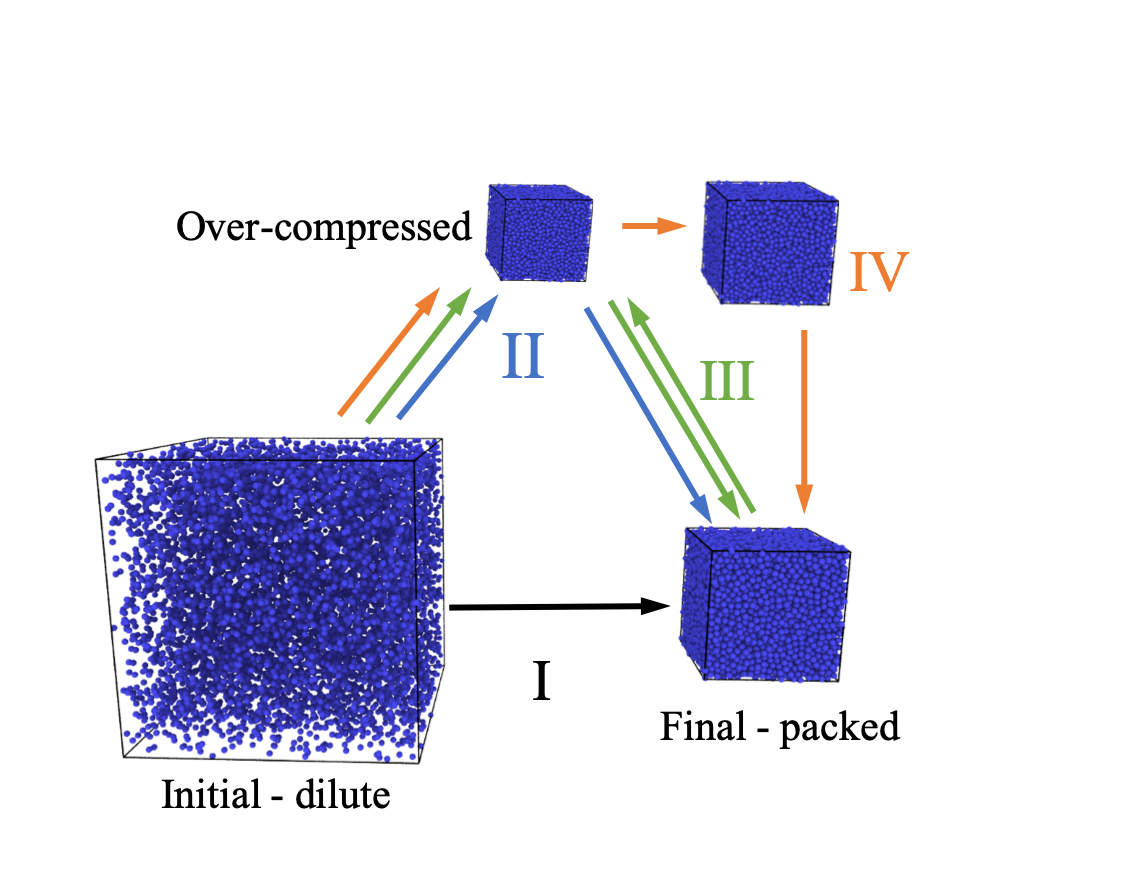}
    \end{center}
    \caption{Schematic of the pressure-controlled isotropic compression methods used in simulations. Procedures are illustrated as black arrows for methods I (black), II (blue), III (green) and IV (orange), and are described in the text.}
    \label{fig:protocols}
\end{figure}

For each pressure, protocol and friction simulated, 6 packings of $N=10^4$ monodisperse particles are generated. Property uncertainties are calculated as the standard deviation from the 6 different packings.
Simulations are initialized with particles at random positions and low volume fraction $\phi_0 = 0.05$. The initial volume fraction $\phi_0$ did not affect the properties of the final packing studied here, so long as $\phi_0$ is well below the jamming volume fraction ($\phi_0 < \phi_{\text{jam}} - 0.3$). 
Initial transnational and rotational velocities were set to zero, except when otherwise noted in which case velocities sample a Gaussian distribution with a mean of 0 and a standard deviation to produce an applied initial kinetic energy.
The simulation time step was set to $\Delta t=0.02\tau$. Timestep $\Delta t=0.002\tau$ was also tested and did not change the results for the pressures studied within the uncertainties. 
After initialization, the particles are isotropically compressed.
Although the precise initial state of the particles did not impact the packings, the path to final state has a large impact. Path dependence is expected for granular particles, because the system is dissipative and far-from-equilibrium. 
To sample the possible methods to pack with a stress-tensor control, particles are compressed using one of the following five methods: (I) starting at $\phi_0 = 0.05$, at $t=0$ a constant pressure $P_{a,f}$ is applied until the system jams, (II) after the system jams at an initial, high pressure $P_{a,0} > P_{a,f}$, the applied pressure is instantaneously decreased to $P_{a,f}$, (III) method II is repeated $N_{\text{cycle}}$ times, where the system jams after each $P_{a,0}$ and $P_{a,f}$ is applied, (IV) after the system jams at an initial, high pressure $P_{a,0} > P_{a,f}$, the applied pressure is step-wise decreased, by a fraction of $P_{a,0} - P_{a,f}$, re-jamming at each step until the system reaches $P_{a,f}$, and (V) is the same as method IV but volume changes, not pressure, similar to a method used in previous simulations~\citep{Silbert2010}. Protocols I-IV are schematically shown in Figure \ref{fig:protocols}. 

Beyond the effect of the packing path and method, the stress tensor can be constrained in different ways. 
Triaxial compression tests are a close experimental equivalent to the simulation constraints on the stress tensor for isotropic compression~\citep{reddyDevelopmentTrueTriaxial1992}. However the presented simulations use periodic boundary conditions instead of walls. 
We simulate three cases of applied symmetric stress tensors $\mathbf{\sigma}_a$: (i) $P_a = \sigma_{a,xx} = \sigma_{a,yy} = \sigma_{a,zz}$ and $\sigma_{a,xy} = \sigma_{a,xz} = \sigma_{a,yz} = 0$, (ii) $P_a = (\sigma_{a,xx}+\sigma_{a,yy}+\sigma_{a,zz})/3$ and $\sigma_{a,xy} = \sigma_{a,xz} = \sigma_{a,yz} = 0$ and (iii) $P_a = \sigma_{a,xx} = \sigma_{a,yy} = \sigma_{a,zz}$, while $\sigma_{a,xy}, \sigma_{a,xz}$ and $\sigma_{a,yz}$ are unspecified and the cell remains rectilinear\footnote[3]{To apply those symmetric stress tensors in LAMMPS~\citep{thompsonLAMMPSFlexibleSimulation2022}, use \texttt{fix nph/sphere} with the following options: (i) \texttt{xy 0 0 1 xz 0 0 1 yz 0 0 1} and (ii) \texttt{xy 0 0 1 xz 0 0 1 yz 0 0 1 couple xyz}. Case (iii) does not need additional options. See LAMMPS documentation for more details.}.
At packing in all these simulations, the final stress tensor equals the applied stress tensor. The differences in the stress tensor of the final packings illustrates the importance of understanding the choice of applied stress tensor.

\begin{figure}[htbp]
\begin{center}	
\includegraphics[width=12cm]{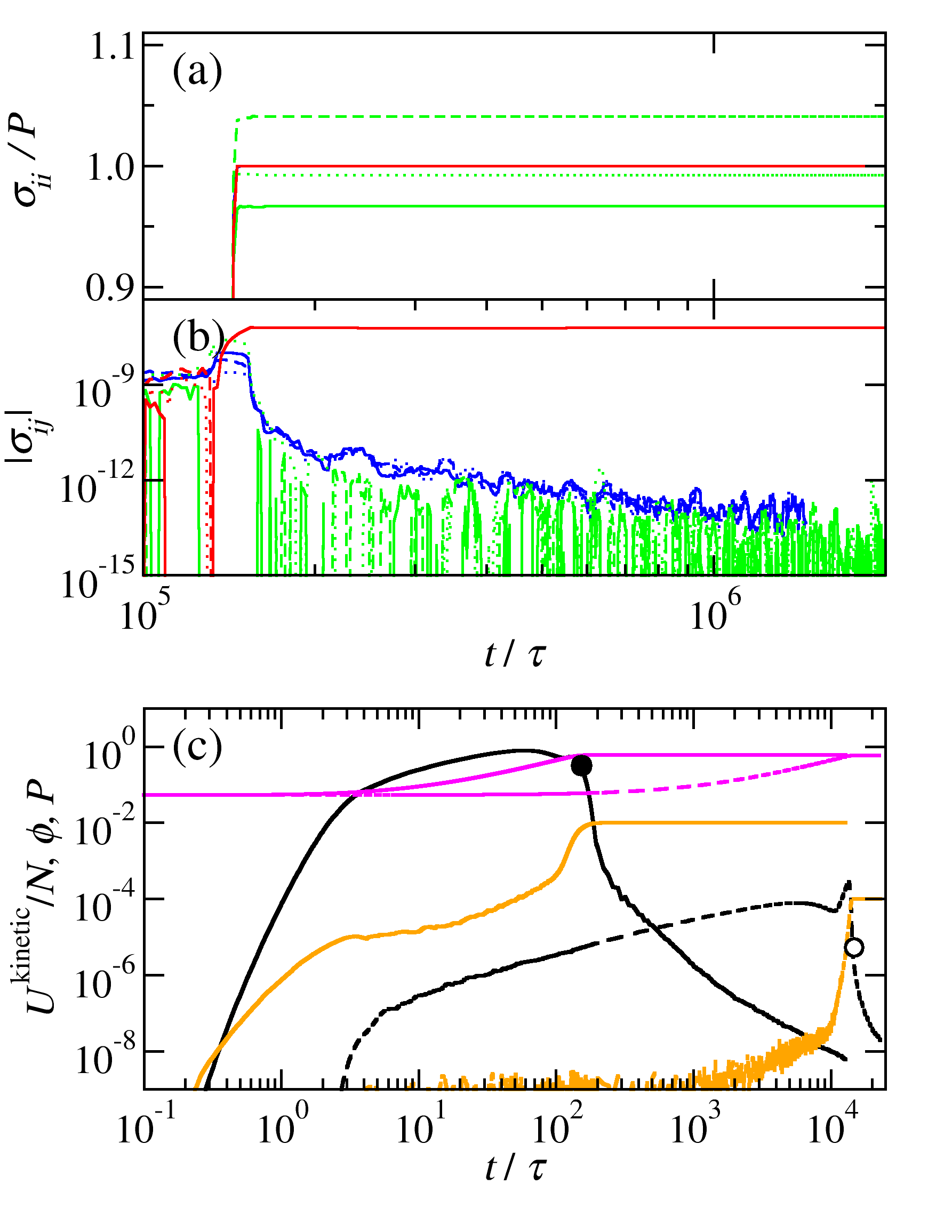}
\end{center}
\caption{The (a) diagonal and (b) off-diagonal components of the applied stress tensor for $P_{a}=10^{-5}$. Three applied stress tensor constraints are plotted: $P_a = \sigma_{a,xx} = \sigma_{a,yy} = \sigma_{a,zz}$, $\sigma_{a,xy} = \sigma_{a,xz} = \sigma_{a,yz} = 0$ (blue), $P_a = (\sigma_{a,xx} + \sigma_{a,yy} + \sigma_{a,zz})/3$, $\sigma_{a,xy} = \sigma_{a,xz} = \sigma_{a,yz} = 0$ (green) and $P_a = \sigma_{a,xx} = \sigma_{a,yy} = \sigma_{a,zz}$, with unspecified values of $\sigma_{a,xy}, \sigma_{a,xz}$ and $\sigma_{a,yz}$ (red) using packing method I. The different components of the stress tensor are plotted as different line types: xx, xy (solid lines), yy, xz (dashed lines) and zz, yz (dotted lines). The off-diagonal components of the stress tensor are shown as averages over 10 timesteps for clarity. The red lines lie on top of the blue lines because they have the same diagonal applied stress components in (a).
	(c) The kinetic energy (black), measured pressure (orange) and the volume fraction (magenta) as a function of time for $P_a=10^{-2}$ (solid lines) and $P_a=10^{-4}$ (dashed lines) using method I. The applied stress tensor is: $P_a = \sigma_{a,xx} = \sigma_{a,yy} = \sigma_{a,zz}$, $\sigma_{a,xy} = \sigma_{a,xz} = \sigma_{a,yz} = 0$. The jamming time $t_{\text{jam}}$, determined as the inflection point of the kinetic energy for $t>P_{\text{damp}}$, is plotted as black circles for $P_a=10^{-2}$ (filled) and $P_a=10^{-4}$ (open). For (a), (b) and (c) the simulation cell parameters are $P_{\text{damp}}=2$ and $f_{\text{drag}}=0.1$, and the friction state is $\mu_s=0.2$.}
\label{fig:stresses}
\end{figure}

All of the stress-tensor constraints form mechanically stable, jammed configuration. 
However, the final stress tensors differ. Figure \ref{fig:stresses}a-b shows the six components of the diagonal and off-diagonal components of the stress tensor, respectively, using method I. The off-diagonal stress components show the largest differences, see Figure \ref{fig:stresses}b.
Simulation cells that are not allowed to tilt, where $\sigma_{a,xy}, \sigma_{a,xz}, \sigma_{a,yz}$ are unspecified, had nonzero, albeit small, values of off-diagonal stress at jamming. 
Those non-zero shear stresses could lead to different yield stresses~\citep{Dagois-Bohy2012}.
Simulation cells that are allowed to tilt, have off-diagonal stress values that decay to zero, and average angles off the orthorombic box of $90 \pm 0.003^{\circ}$, for all frictions and pressures tested. 
The diagonal components of stress $\sigma_{a,xx}$, $\sigma_{a,yy}$ and $\sigma_{a,zz}$ are less affected by the constraints. 
Unless noted otherwise, simulations in Sec. \ref{sec:results} set diagonal members of the applied stress tensor to the pressure, $P_a = \sigma_{a,xx} = \sigma_{a,yy} = \sigma_{a,zz}$, and off-diagonal members to zero, $\sigma_{a,xy} = \sigma_{a,xz} = \sigma_{a,yz} = 0$. 
Such precise control on stress is usually unattainable for experimental packing schemes. However the differences in final states demonstrate the importance of knowing the relevant stress and volume controls in experimental and simulation protocols.

Using method I and the stress constraint defined as case i, a representative simulation time progression of the kinetic energy, volume fraction and pressure are shown in Figure \ref{fig:stresses}c. At $t=0$ the kinetic energy and pressure are zero, except for cases with defined initial pressure discussed in Sec.~\ref{sec:explanation}, at the initial volume fraction $\phi = 0.05$. As the simulation cell volume decreases and picks up momentum, the particle velocities increase due to affine motion, and the kinetic energy and pressure increase. At $t \simeq P_{\text{damp}}$ the Parinello-Rahman algorithm starts to control the pressure and the simulation cell momentum, and the kinetic energy decreases. Near jamming, the kinetic energy decreases by several orders of magnitude and the volume fraction plateaus. The pressure jumps to the applied value as contacts form, with the full applied stress tensor satisfied by the constraints. The near-jamming behavior was similar for all systems studied. However, there are differences at earlier time based on the barostat parameters and initial configuration. Lower values of drag approach the applied pressure faster but with more oscillations.

The volume fraction $\phi$ and coordination number $Z$ are the key parameters calculated in this study. 
Both $\phi$ and $Z$ are calculated without ``rattlers'', particles that have too few contacts to contribute to the mechanical stability of the packings. Rattlers are identified if $Z_i < 6$ frictional ($\mu_s>0.01$) and $Z_i < 4$ for frictionless particles, where $Z_i$ is the number of contacts of particle $i$.
The critical friction value $\mu_s=0.01$ was chosen because it is the point where friction has an appreciable impact on $\phi$ and $Z$~\citep{santosGranularPackingsSliding2020}. 
Rattlers are identified iteratively, so that the number of contacts per particle decreases based on the number of rattlers in contact with the particle. If the number of contacts decrease enough to constitute a rattler, by removing neighboring rattlers, it is counted as such.

All of the packings generated were taken from the final simulation configuration, after the simulation was run for at least twice the jamming time. The time to jam depends on the method, the particle and barostat parameters, and therefore some simulations ran longer than others. The inflection point of the kinetic energy, plotted as symbols in Figure \ref{fig:stresses}c, corresponds well with the point where volume fraction stops changing and is a good estimate of the time to jam. However, the volume fraction is not strictly constant once the simulation cell stops moving, and increases slowly for some longer time. To allow for these changes, we run to $t/\tau = 10^6$ for $P_a = 10^{-4}, f_{\text{drag}} = 0.0$ and $P_{\text{damp}} = 2.25$ which is well above the time to jam $t_{jam}\sim 1.5$x$10^{4} \tau$. 
The inflection point in kinetic energy defines $t_{\text{jam}}$. The time to jam is proportional to the applied pressure, $t_{\text{jam}} \propto \frac{f_{\text{drag}}}{P_aP_{\text{damp}}}$, and thus the simulation time was scaled accordingly for lower $P$ and/or higher $f_{\text{drag}}$.

\section{Results} \label{sec:results}

\subsection{Packing method dependence}\label{sec:protocols}
\begin{figure}[htbp]
\begin{center}	
\includegraphics[width=15cm]{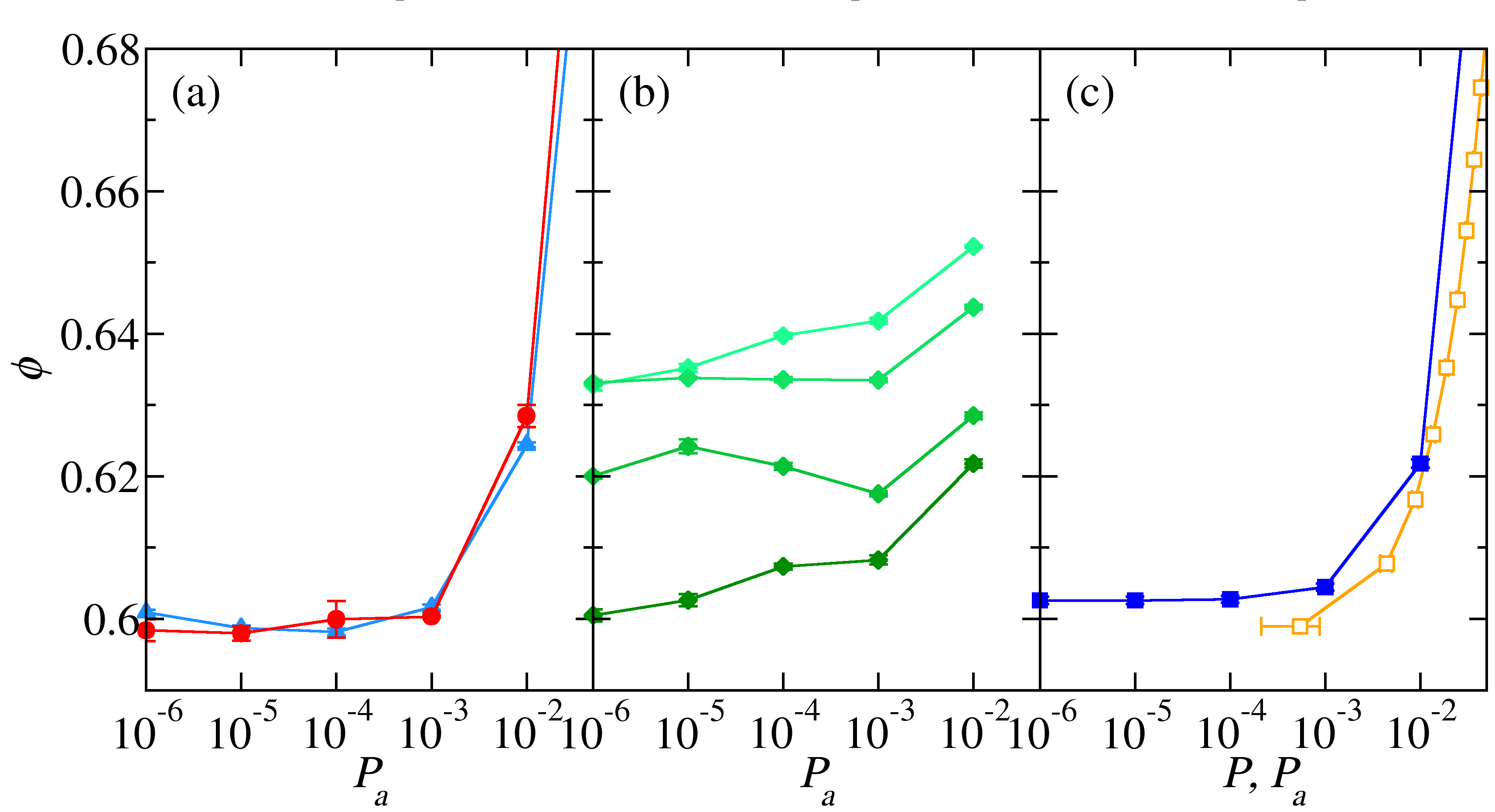}
\end{center}
\caption{(a) Method I (red circles) packing fraction as a function of pressure $\phi(P_a)$ is compared with method II (light blue triangles), where $P_{a,0}=10^{-4}$ and $P_{a,f}=P_a$. (b) Method III, akin to tapping, is shown after different number of compressions $N_{\text{compress}} =$ 1 (dark green diamonds), 10, 100 and 1000 (light green diamonds). The packings are compressed to $P_{a,0}=10^{-1}$ in between relaxations. (c) Progressive compression methods with stress- (IV, blue squares) and volume- (V, orange squares) control show different ranges of pressure. Method V volume step changes were constant $\Delta\phi=0.01$. Particles are frictional $\mu_s = 0.2$, and are packed with simulation cell parameters $P_{\text{damp}}=2.25$ and $f_{\text{drag}}=0$. Uncertainties are similar to the symbol size.}
\label{fig:methods}
\end{figure}

To explore different routes for frictional particle packing~\citep{Silbert2002,Shundyak2007,Silbert2010} we applied various isotropic compression methods to particles with sliding friction. In this subsection the packings were formed at different applied pressures $P_a$, where the internal pressure of the mechanically stable packing $P_{\text{int}}=P_a$, with sliding friction $\mu_s=0.2$. The packing volume fraction is between the frictionless and high friction limits at $\mu_s=0.2$, where $\mu_s=0.2$ is in the middle of experimentally observed material friction range~\citep{Farrell2010}. 
The low-pressure range can be jammed stably, at low computational cost. 
The packing behavior generated by pressure-controlled compression methods I-V are shown in Figure~\ref{fig:methods} and detailed in Sec.~\ref{sec:packingmethod}.

Figure~\ref{fig:methods}(a) shows methods I and II, under- and over-comperssion. Method I applies a pressure at $t=0$ to a dilute packing; a lower pressure translates to slower compression.
Method II follows method I at first, where a initial pressure is applied $P_{a,0}$ to a dilute system ($\phi=0.05$) to form a mechanically stable packing. A lower pressure $P_{a,f}$ is applied to the packing formed at $P_{a,0}$ to form a new mechanically stable packing. The pressure on the x-axis of the left panel of Figure~\ref{fig:methods} is the $P_{a,f}$ for method II. 
The Supplementary Information includes method II packing fractions with other initial pressures $P_{a,0}$.
As expected~\citep{OHern2003,Silbert2010}, $\phi$ from method I, decreases monotonically. Although the absolute values between methods I and II are similar, method II has a minimum with pressure. The non-monotonic pressure dependence is analyzed in Sec.~\ref{sec:explanation}.

\begin{figure}[htbp]
\begin{center}	
\includegraphics[width=16cm]{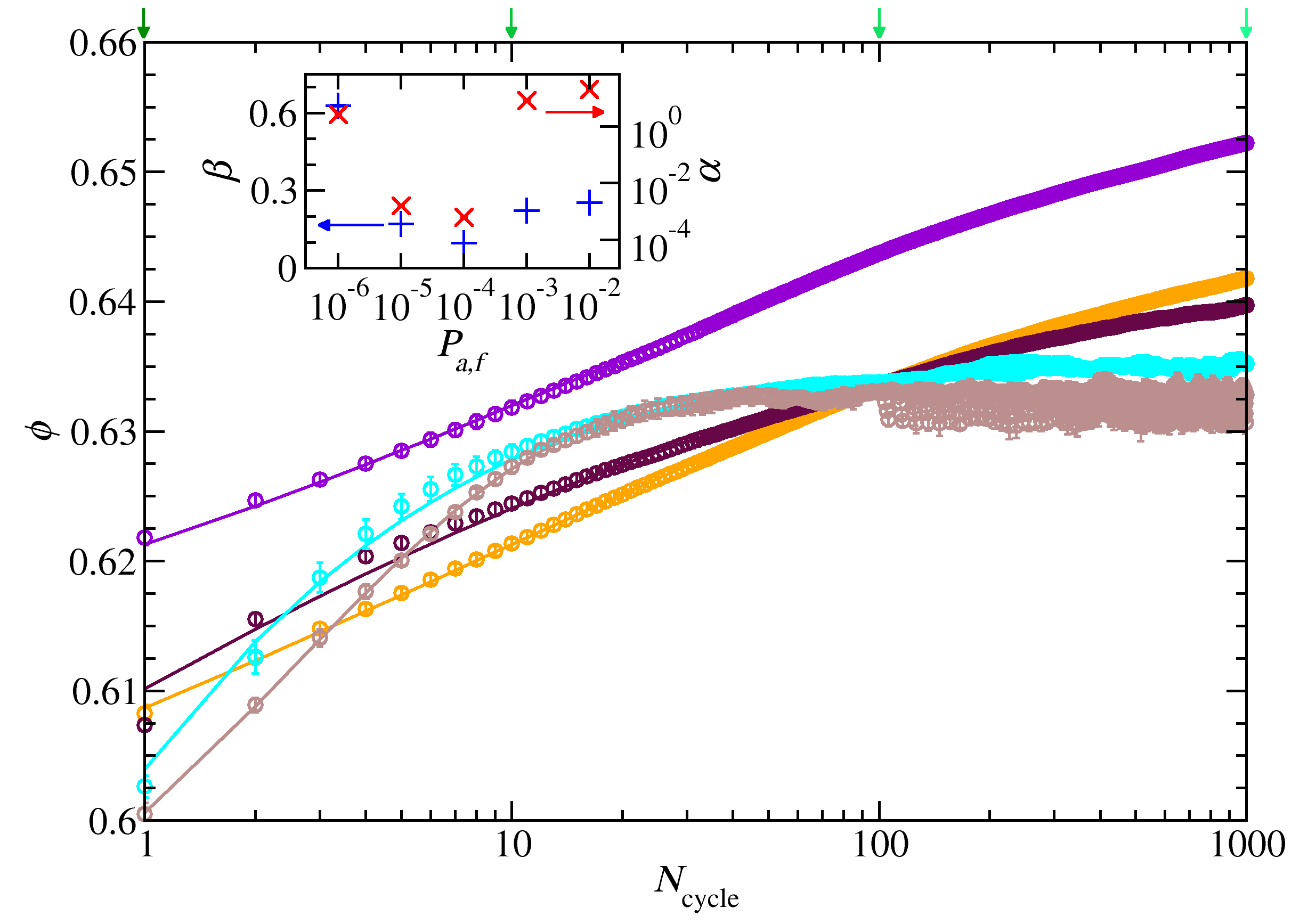}
\end{center}
\caption{Volume fraction $\phi$ increases monotonically with $N_{\text{cycle}}$ using Method III, by cycling from $P_{a,0}=10^{-1}$ to different low-pressure compression values $P_{a,f} = 10^{-2}$ (magenta), $10^{-3}$ (orange), $10^{-4}$ (maroon), $10^{-5}$ (cyan) and $10^{-6}$ (brown). Lines drawn are stretched-exponential fits to simulation data. The KWW fit parameters $\alpha$ (red crosses, left inset axis) and $\beta$ (blue pluses, right inset axis) are plotted in the inset as a function of the applied pressure $P_a$. The arrow colors at the top of the graph indicate the number of compressions that correspond with the $\phi(P)$ data shown in Figure~\ref{fig:methods}(b). Uncertainties are similar to the symbol size.}
\label{fig:cycles}
\end{figure}

Like method II, method III can lead to monotonic or non-monotonic $\phi(P_a)$. Method III, essentially, cyclically repeats method II. The first cycle in method III, $N_{\text{cycle}}=1$, is the same as method II with the $P_{a,0}=10^{-1}$, at which point there is no minimum in $\phi(P_a)$, shown in Figure~\ref{fig:methods}b. The minimum in $\phi(P_a)$ appears after a few cycles ($5<N_{\text{cycle}}<100$) and disappears at higher cycles ($N_{\text{cycle}}>100$). 

The non-monotonic $\phi(P_a)$ behavior, seen in Figure~\ref{fig:methods}b, occurs over a range of $N_{\text{cycle}}$, shown in Figure~\ref{fig:cycles}. For each $P_a$, $\phi(P_a,N_{\text{cycle}})$ increase monotonically with $N_{\text{cycle}}$. 
Lower pressures $P_{a} < 10^{-4}$, compact at a faster rate with respect to $N_{\text{cycle}}$ and saturate as $N_{\text{cycle}}\to \infty$. 
The lower $P_a$ packing fractions crossing the higher $P_a$ values, around $N_{\text{cycle}}=5$ and $N_{\text{cycle}}=70$, is the same result as the non-monotonicity observed in $\phi(P_a)$, see Figure~\ref{fig:methods}b. 
Yet, since the lower $P_a$ packings compaction asymptotes at fewer $N_{\text{cycle}}$, higher $P_a$ packings are denser, and $\phi(P_a)$ is monotonic at higher $N_{\text{cycle}}$. The lower pressures have a larger difference with $P_{a,0}$, which allows more time to pack and re-form contacts to build more compact networks with fewer $N_{\text{cycle}}$.
At high $N_{\text{cycle}}$ method III forms denser packings with more predictable monotonic $\phi(P_a)$ behavior. 

The behavior observed in the $\phi(N_{\text{cycle}})$ are captured by fits to a Kohlrausch-Williams-Watts (KWW) law ~\citep{Kohlrausch1854,Williams1970}:
\begin{equation}\label{eqn:KWW}
\phi(N_{\text{cycle}})=\phi_\infty - (\phi_\infty-\phi_0)e^{-\left(N_{\text{cycle}}/\alpha\right)^\beta}
\end{equation}
where the fitting parameters are $\phi_\infty$, $\phi_0$, $\alpha$ and $\beta$. The intercept $\phi_0$ and asymptote $\phi_\infty$ values are monotonic, inferred by the low and high  $N_{\text{cycle}}$ curve values in Figure~\ref{fig:cycles}. The Figure~\ref{fig:cycles} inset shows that the parameters $\alpha$ and $\beta$ are nonmonotonic with pressure.   
The KWW fit parameters $\alpha$ and $\beta$ quantify the trends in $\phi(N_{\text{cycle}},P_a)$ and show different behavior above and below $P_a=10^{-4}$. 

The KWW and a logarithmic heuristic~\citep{Knight1995} fits have been applied to experimentally tapped packings. 
The KWW fit had consistently lower residual standard deviations, compared to logarithmic heuristic fit for the presented data, as seen by~\citep{Richard2005}. 
Method III is considerably different from the experimental tapping protocols~\citep{Knight1995,Philippe2002}, which are compressed in all directions, have no walls and and vary the peak tap acceleration, not the pressure, and lead to denser volume fractions $\phi>0.64$. 
KWW fits to experimental data~\citep{Knight1995,Philippe2002} parameters range from $1 < \alpha < 500$ and $0.14 < \beta < 0.65$. Simulation and experimental exponential KWW fit parameter $\beta$ are in the same range. The $\alpha$ fit parameters have a different meaning in experiments, which track $\phi(t)$ not $\phi(N_{\text{cycle}})$, in which case $\alpha$ is a rate.
However both experiments and simulations found that $\beta$ increase and $\alpha$ decreases with increasing packing intensity.
However the DEM simulations showed that, like experimental tapping, ``loose'' packings compact with tapping~\citep{Knight1995,Rosato2010}. ~\citet{Kumar2016} observed similar behavior and found that memory of the deformation theory could explain the denser-than-experiments volume fractions.

Methods IV and V, shown in Figure~\ref{fig:methods}c, differ from method II by gradually, instead of instantaneously, decreasing the applied, target pressure, at each step allowing the particles to pack after dilation. Method IV uses pressure-controlled compression, like in methods I-III, and in method V the volume is decreased by $\Delta\phi = 0.01$. Smaller volumetric decreases can lead to looser packings~\citep{Silbert2010}. Neither method IV or V has a minimum in $\phi(P_a)$, as observed in method II. The absence of a minimum is likely because the volume change is not large enough to break-up the majority of the contact network. Stable packings could not be formed with method V for $\phi<0.599$ and $P<5\text{x}10^{-4}$. ~\citet{Silbert2010} observed similar volume-controlled packing limits. 
Ramped-pressure compression simulations of cohesive, frictional grains have exhibied strong history and protocol dependence~\citep{nan_ultraslow_2023}.
These methods show that stable packings of the same model frictional particles with the same stress state can have a wide range of volume fractions, and are path dependent.

\subsection{Non-monotonic volume fraction-pressure dependence}\label{sec:explanation}

Depending on the packing protocol the final volume fraction is not always a monotonically decreasing function of pressure. 
The minimum in $\phi(P_a)$ shown in Figure~\ref{fig:methods}a-b for packing methods II and III showcases the protocol-dependent nature of the packing process. 
A minimum is not observed in the coordination number, which is relatively insensitive to packing protocol.
This leads to the possibility of two packings with the same volume fraction, but different coordination numbers.
The initial kinetic energy, drag coefficient and friction are varied to observe the scale protocol parameter impacts on the non-monotonic behavior.

\begin{figure}[htbp]
\begin{center}	
\includegraphics[width=9cm]{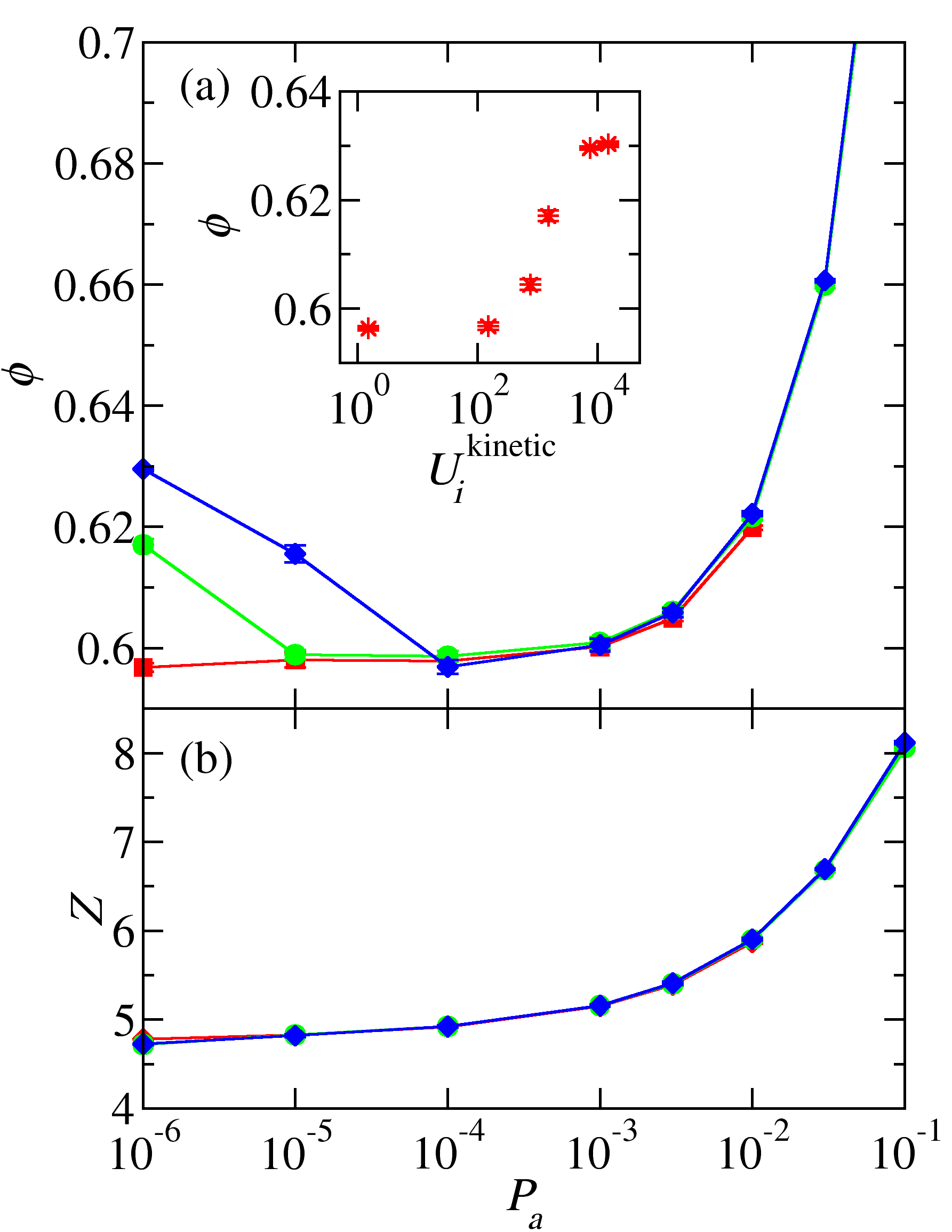}
\end{center}
\caption{(a) Packing fraction $\phi$ and (b) average coordination number $Z$ without rattlers as a function of the pressure $P_a$. Packings were generated with method I, $P_{\text{damp}}=2.25$, $f_{\text{drag}}=0$ and varied amounts of initial total translational kinetic energy $U_i^{\text{kinetic}}=$ 0 (red squares), $1.5\times10^4$ (greeen circles) and $7.5\times 10^{4}$. (inset) Volume fractions as a function of the initial total translational kinetic energy $U_i^{\text{kinetic}}$ at low pressures asymptote to the zero and high pressure values. Coordination number symbols lie on top of each other. Uncertainties are similar to the symbol size.}
\label{fig:PhiZPress}
\end{figure}

The initial pressure and kinetic energy are important contributions to the packing microstructure.
Packings in Figures~\ref{fig:methods} and \ref{fig:cycles} were initiated with zero initial kinetic energy and pressure.
Increasing the average initial particle translational kinetic energy causes a volume fraction minimum using packing method I.  
Figure~\ref{fig:PhiZPress}a shows the role of initial kinetic energy $U_i^{\text{kinetic}}$. The $\phi(P_a)$ minimum is more pronounced with increasing $U_i^{\text{kinetic}}$. 
Figure~\ref{fig:PhiZPress}b demonstrates that packings with the same particle interactions can be made with the same volume fraction, for example $\phi=0.62$, with an average one fewer contact per particle (compare $U^{\text{kinetic}}_{i}=1.5\times10^{-4}$ at $P_a=10^{-6}$ and $10^{-2}$ in Figure~\ref{fig:PhiZPress}).
The Figure~\ref{fig:PhiZPress}a inset shows that the increases the initial kinetic energy $U^{\text{kinetic}}_{i}$ increases the depth of the $\phi(P_a)$ minimum, but has a limit of about $\Delta\phi=0.03$.
The Supplementary Information shows the role of initial kinetic energy on the transient approach to packing and on method II packings.
 
The minimum value of $\phi$ in Figure~\ref{fig:PhiZPress}a occurs at $P_a = 10^{-4}$, comparable to the lowest pressures (for intermediate to high $\mu_s$) accessible in volume-controlled studies (see Figure~\ref{fig:methods}c and references~\citep{Shundyak2007,Silbert2010}). 
The behavior of the cyclical packings, generated with Method III, also transition at  $P_a=10^{-4}$, specifically the KWW fit parameters $\alpha$ and $\beta$ in the Figure~\ref{fig:cycles} inset.

\begin{figure}[htbp]
\begin{center}
\includegraphics[width=12cm]{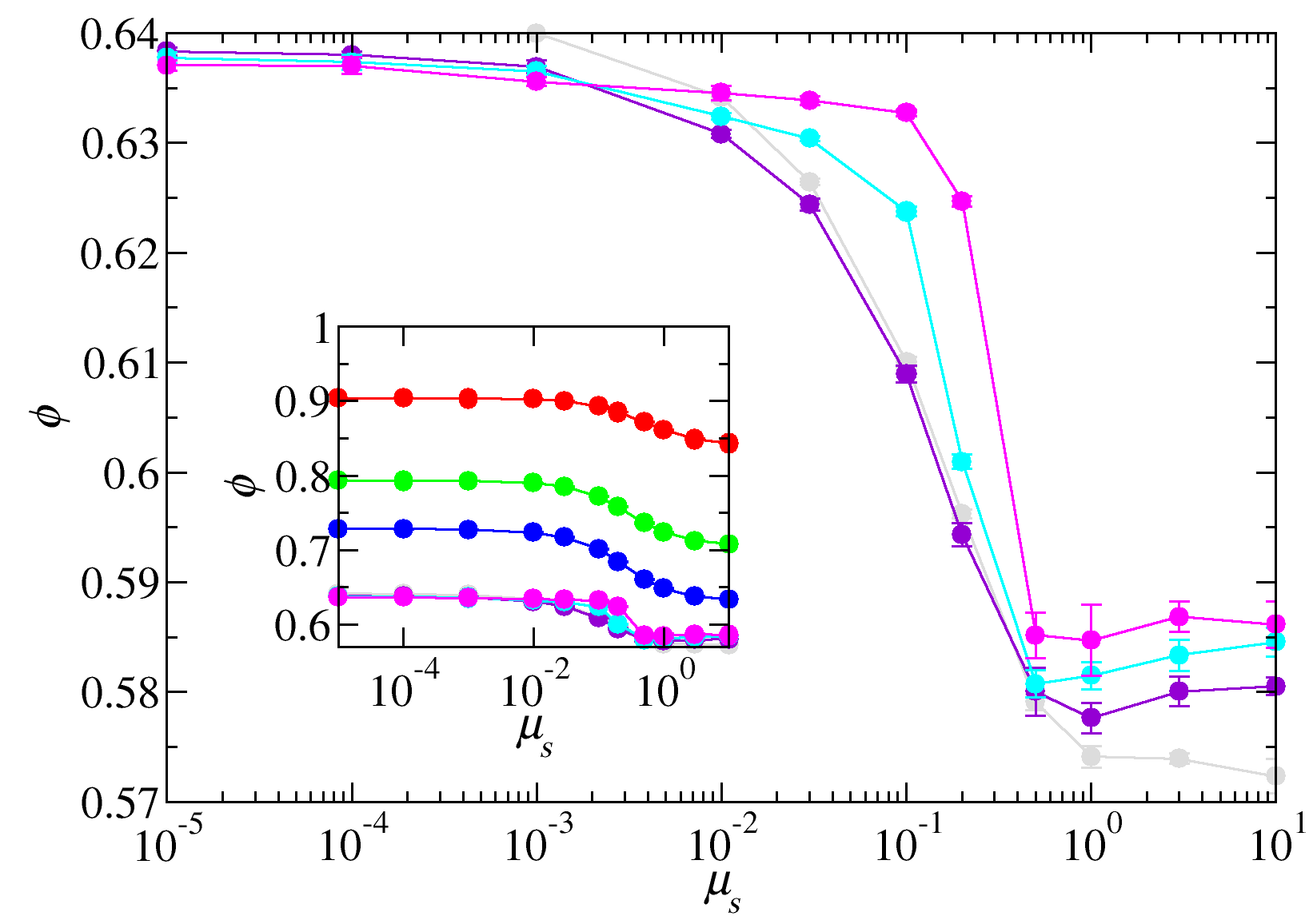}
\end{center}
\caption{Packing fraction $\phi$ as a function of sliding friction coefficient $\mu_s$ near the minimum in Figure \ref{fig:PhiZPress} for different pressures, $P_a$ = $2$x$10^{-1}$ (red), $1$x$10^{-1}$ (green), $5$x$10^{-2}$ (blue), $P_a = 10^{-3}$ (grey), $10^{-4}$ (purple), $10^{-5}$ (cyan) and $10^{-6}$ (magenta). The inset shows a larger range of $\phi$ as a function of sliding friction coefficient $\mu_s$. These packings were generated using method I with an initial pressure and protocol parameters $P_{\text{damp}}=2$, $f_{\text{drag}}=0.1$. Uncertainties are similar to the symbol size.}
\label{fig:Phimus}
\end{figure}

Particle friction is known to lower packing fraction and coordination number, but also changes the $\phi(P_a)$ minima. 
Packing fractions in Figures~\ref{fig:methods}-\ref{fig:PhiZPress} are from particles with intermediate friction $\mu_s=0.2$. 
The non-monotonicity in $\phi(P_a)$ effects the friction dependence of $\phi(\mu_s)$ as shown in Figure~\ref{fig:Phimus}.
The general form of $\phi(\mu_s)$ is similar to previous studies of packing with sliding friction~\citep{Shundyak2007,santosGranularPackingsSliding2020}, however the initial pressure and drag changes the pressure dependence. 
For larger pressures, $P_a>10^{-3}$, the $\phi(\mu_s)$ shape remains the same.
For $P_a<10^{-3}$, frictionless particles approach the hard-sphere limit and $\phi$ approaches the $\mu_s=0$ maximally jammed state. 
The non-monotonicity with pressure occurs for frictions $\mu_s>10^{-3}$, where the different pressure curves cross.
Lowering the pressure narrows the low-to-high $\mu_s$ transition, when initialized with non-zero pressure.
Although it seems that $\phi(\mu_s)$ tends to a step function as $P_a \to 0$, that behavior depends on protocol. Going to lower pressures to see if a step function arises is computationally difficult because the time to jam the system scales inversely with the applied pressure. 
The $\phi(\mu_s)$ behavior, as does $\phi(P_a)$, highlights the interdependence of particle interaction and control parameters. 

\begin{figure}[htbp]
\begin{center}	
\includegraphics[width=12cm]{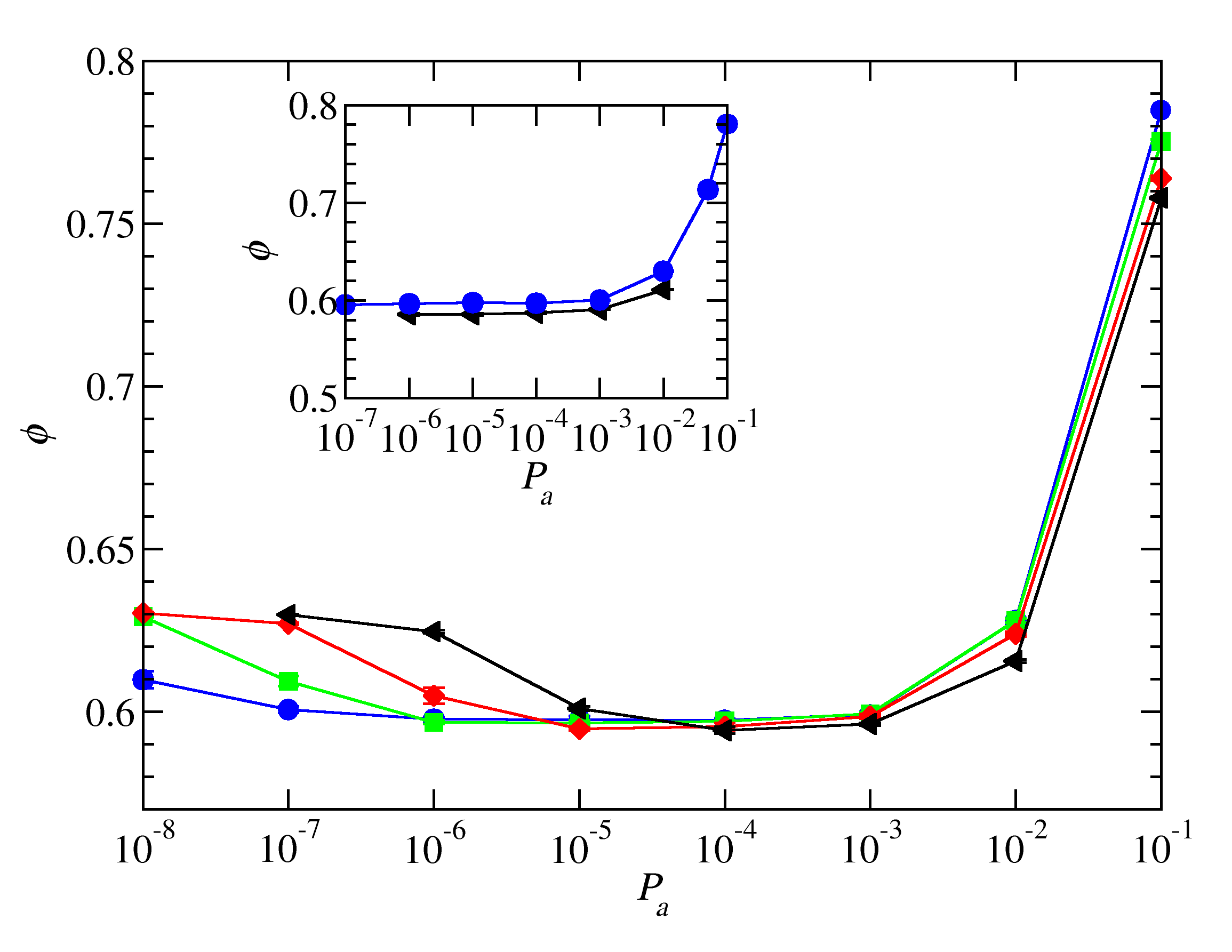}
\end{center}
\caption{Packing fraction as a function of the pressure $\phi(P_a)$ with particle friction $\mu_s = 0.2$. Initial kinetic energy from overlap $\langle P_i^{\text{kinetic}}\rangle=9.5\times10^{-4}$ causes a drag factor-dependent non-monontonic behavior (inset) No initial kinetic energy or pressure $P_i^{\text{kinetic}}=0$ yields monotonic decreasing $\phi$ with decreasing $P$. 
Drag is applied to the simulation cell by different drag factors $f_{\text{drag}}$ = 0.0 (blue circles), 0.1 (green squares), 0.3 (red diamonds) and 1.0 (black triangles) with $P_{\text{damp}}=2.25$. Packings were generated with method I, and the $\sigma_{a,xx} = \sigma_{a,yy} = \sigma_{a,zz} = P_a$, $\sigma_{a,xy} = \sigma_{a,xz} = \sigma_{a,yz} = 0$ stress-tensor constraint. Uncertainties are similar to the symbol size.}
\label{fig:drag}
\end{figure}

To model packing of particles in the presence of a viscous fluid, we include a drag term $f_{\text{drag}}$. Like the initial pressure, the introduction of a drag can have significant affect on the final packing fraction.
Figure~\ref{fig:Phimus} shows data for packings generated with drag, while packings in Figures ~\ref{fig:methods}-\ref{fig:cycles} have no drag. 
Figure \ref{fig:drag} shows that although drag can change the volume fraction, a minimum in $\phi(P_a)$ is present for all values of $f_{\text{drag}}$ packed using method I with non-zero initial pressure. 
For lower pressure, $P_a<10^{-4}$, the $\phi(P_a)$ minimum is more narrow for larger drag $f_{\text{drag}}$. A limiting value of $\phi(P_a \to 0) \simeq 0.63$ is the same with all simulation cell drags. 
The inset in Figure~\ref{fig:drag} shows that drag has a small effect on packings when initialized with zero pressure. 
The $\phi(P_a)$ dependence on $f_{\text{drag}}$ demonstrates another of many components of protocol design that impact the final packing of frictional particles.

\begin{figure}[htbp]
\begin{center}
\includegraphics[width=16cm]{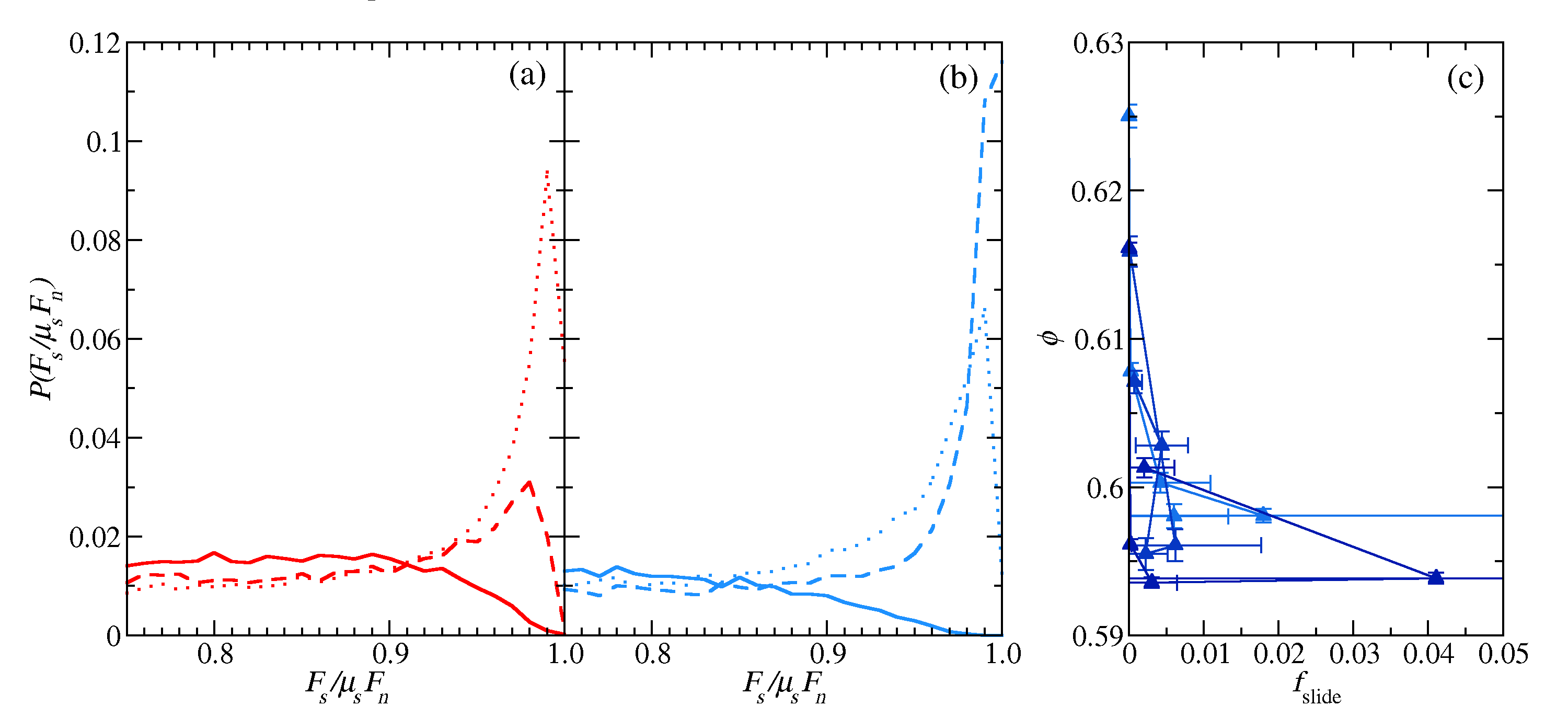}
\end{center}
\caption{Probability distribution of the sliding force normalized by the maximum, $\mu_sF_n$, for different pressures: $P_a =$ 10$^{-2}$ (solid lines), 10$^{-4}$ (dotted lines) and 10$^{-6}$ (dashed lines). Method I (a, red) and method II ($P_{a,0}=10^0$) (b, blue) show different $\phi(P_a)$ behavior, see Figure \ref{fig:methods}. 
(c) The volume fraction as a function of the fraction of contacts at the sliding friction criteria $\mu_s F_s = F_n$. The packings were generated with an initial pressure and protocol parameters $P_{\text{damp}}=2$, $f_{\text{drag}}=0.1$. Large uncertainties in $f_{\text{slide}}$ are due to small absolute denominator values.}
\label{fig:PressureForce}
\end{figure}

The distribution of forces offers an explanation for the non-monotonicity of volume fraction with pressure. 
The distributions of sliding forces, normalized by their maximum $\mu_s F_n$, are shown in Figure~\ref{fig:PressureForce} for methods I and II. Both have non-zero initial pressure; method I has the non-monotonic $\phi(P_a)$ and method II does not. 
The probability distribution is normalized so that $\Sigma_{F_s/\mu_sF_n}{P}(F_s/\mu_sF_n) = 1$. 
The impact of the $\phi(P_a)$ non-monotonicity is visible in ${P}(F_s/\mu_sF_n)$ for $P_a\le10^{-4}$. For method I, contacts near the Coulomb criteria $(F_s/\mu_sF_n>0.94)$ become less likely as pressure decreases from $P_a=10^{-4}$ to $P_a=10^{-6}$, Figure~\ref{fig:PressureForce}a. For method II, which does not show non-monotonicity in $\phi(P_a)$, contacts are more likely to be near the Coulomb criteria as the pressure decreases, Figure~\ref{fig:PressureForce}b. Method II shows the more expected behavior because $P_a \propto F_n$. 

The peak location of ${P}(F_s/\mu_sF_n)$ is another manifestation of the non-monotonic $\phi(P_a)$ behavior. The ${P}(F_s/\mu_sF_n)$ peak is shifted below $F_s/\mu_sF_n=1$ for $P_a \le 10^{-4}$ in method I. This implies that those larger sliding forces were able to relax, due to slower compression. And as the sliding friction contacts weaken, the contacts become less frictional. Seemingly, the tangential constraint sets the average coordination number regardless of its strength. Therefore, the sliding constraint network is maintained as the constraint weakens, but the packing is able to compact.
Based on this hypothesis, one would expect the volume fraction to be monotonic not with pressure, but with the number of sliding contacts. The fraction of sliding contacts $f_{\text{slide}}$, where $\mu_s F_s = F_n$, also has a non-monotonic dependence with pressure. The $\phi(f_{\text{slide}})$ dependence for method II is shown in Figure \ref{fig:PressureForce}c. The fraction of contacts at the Coulomb criteria has an inverse relationship with volume fraction, which yields a monotonic $\phi(f_{\text{slide}})$ relationship, within uncertainty.
Based on this discussion the packing microstructure depends on the connectivity of the tangential force network, which sets $Z$, but the strength of those tangential contacts, specifically the fraction of sliding contacts, sets $\phi$.

\section{Conclusion}\label{sec:conclusion}
Simulations of 3-dimensional frictional granular particles were packed into mechanically stable configurations were performed by using novel pressure-controlled protocols with various protocol parameters. The protocols modeled bulk-like packings, with periodic boundary conditions and precisely defining internal states of stress. Five packing protocols were studied including: (I) slow compression from a dilute state, (II) slow expansion from a dense state, (III) repetitive compressions and expansions, (IV) pressure-controlled progressive de-compression from a dense state and (V) volume-controlled progressive de-compression from a dense state.

Non-monotonic packing fraction dependence on pressure was observed in multiple methods. This led to configurations packed with the same contact mechanics and the same packing fraction, but up to one average contact less per particle. If dilute initial particle configurations were initialized with non-zero velocities or pressure, the packing fraction has a minimum, whereas the coordination number is monotonic, for the undercompressed protocol (method I). The larger the initial kinetic energy, the larger the minimum packing fraction depth.
The packing fraction minimum with pressure depended on friction and simulation cell drag. More drag led to a more narrow minimum, and the minimum was most pronounced at intermediate frictions.
For the cyclical protocol, method III, non-monotonic packing fraction pressure dependence occurred for intermediate number of packing cycles. The volume fraction evolution with the number of cycles $(\phi(N_{\text{cycles}})$ changed qualitatively with pressure. The parameters for fits to $(\phi(N_{\text{cycles}})$ transitioned at intermediate pressure, changing the low and $N_{\text{cycles}}$ behavior. 

The fraction of frictional contacts were calculated for various packings. We propose that lower volume fractions are supported by a higher fraction of frictional sliding contacts.
The role of friction and pressure on the packing fraction of method I built packings showed that these behaviors disappear for low but significant enough frictions $\mu_s<10^{-2}$. The volume fraction is less sensitive to friction as pressure decreases, indicated by a sharper transition with respect to friction coefficient from frictionless to high frictional behavior.
Further analysis of the contact network properties, possibly with the dynamical matrix and fabric tensor, may better explain the existence of states with high volume fractions and low coordination numbers. 

Stress-controlled packing has a relatively low computational cost and can model bulk-like behavior under various protocols. The volume-controlled protocols are restricted to smaller ranges of pressure than stress-controlled protocols because the precise applied stress-tensor can be controlled.  
The stress-controlled methods can simulate compression paths not studied here to compare to other experimental protocols. 
The work presented here on material behavior along the path of these processes can offer routes to study material- and process-specific packings with simulations.                       

\section*{Conflicts of interest}
There are no conflicts to declare.
\section*{Acknowledgements}
A.S. acknowledges this work was supported, in part, by funding from the NASA Game Changing Development Program. 
I.S. acknowledges support from the U.S. Department of Energy (DOE), Office of Science, Office of Advanced Scientific Computing Research, Applied Mathematics Program under Contract No. DE-AC02-05CH11231.
This work was performed, in part, at the Center for Integrated Nanotechnologies, an Office of Science User Facility operated for the U.S. Department of Energy (DOE) Office of Science. 
Sandia National Laboratories is a multi-mission laboratory managed and operated by National Technology and Engineering Solutions of Sandia, LLC., a wholly owned subsidiary of Honeywell International, Inc., for the U.S. DOE’s National Nuclear Security Administration under contract DE-NA-0003525. 
The views expressed in the article do not necessarily represent the views of the U.S. DOE or the United States Government.

\bibliography{GranularPackingProtocols} 

\providecommand*{\mcitethebibliography}{\thebibliography}
\csname @ifundefined\endcsname{endmcitethebibliography}
{\let\endmcitethebibliography\endthebibliography}{}
\begin{mcitethebibliography}{41}
\providecommand*{\natexlab}[1]{#1}
\providecommand*{\mciteSetBstSublistMode}[1]{}
\providecommand*{\mciteSetBstMaxWidthForm}[2]{}
\providecommand*{\mciteBstWouldAddEndPuncttrue}
  {\def\EndOfBibitem{\unskip.}}
\providecommand*{\mciteBstWouldAddEndPunctfalse}
  {\let\EndOfBibitem\relax}
\providecommand*{\mciteSetBstMidEndSepPunct}[3]{}
\providecommand*{\mciteSetBstSublistLabelBeginEnd}[3]{}
\providecommand*{\EndOfBibitem}{}
\mciteSetBstSublistMode{f}
\mciteSetBstMaxWidthForm{subitem}
{(\emph{\alph{mcitesubitemcount}})}
\mciteSetBstSublistLabelBeginEnd{\mcitemaxwidthsubitemform\space}
{\relax}{\relax}

\bibitem[Snow \emph{et~al.}(2019)Snow, Martukanitz, and Joshi]{Snow2019}
Z.~Snow, R.~Martukanitz and S.~Joshi, \emph{Additive Manufacturing}, 2019, \textbf{28}, 78--86\relax
\mciteBstWouldAddEndPuncttrue
\mciteSetBstMidEndSepPunct{\mcitedefaultmidpunct}
{\mcitedefaultendpunct}{\mcitedefaultseppunct}\relax
\EndOfBibitem
\bibitem[Wischeropp \emph{et~al.}(2019)Wischeropp, Emmelmann, Brandt, and Pateras]{Wischeropp2019}
T.~M. Wischeropp, C.~Emmelmann, M.~Brandt and A.~Pateras, \emph{Additive Manufacturing}, 2019, \textbf{28}, 176--183\relax
\mciteBstWouldAddEndPuncttrue
\mciteSetBstMidEndSepPunct{\mcitedefaultmidpunct}
{\mcitedefaultendpunct}{\mcitedefaultseppunct}\relax
\EndOfBibitem
\bibitem[Gerhard and Reich(2000)]{Gerhard2000}
M.~Gerhard and M.~Reich, \emph{International Review of Hydrobiology}, 2000, \textbf{85}, 123--137\relax
\mciteBstWouldAddEndPuncttrue
\mciteSetBstMidEndSepPunct{\mcitedefaultmidpunct}
{\mcitedefaultendpunct}{\mcitedefaultseppunct}\relax
\EndOfBibitem
\bibitem[Melville and Sutherland(1988)]{Melville1988}
B.~W. Melville and A.~J. Sutherland, \emph{J. Hydraul. Egn.}, 1988, \textbf{114}, 733--9429\relax
\mciteBstWouldAddEndPuncttrue
\mciteSetBstMidEndSepPunct{\mcitedefaultmidpunct}
{\mcitedefaultendpunct}{\mcitedefaultseppunct}\relax
\EndOfBibitem
\bibitem[Torquato \emph{et~al.}(2000)Torquato, Truskett, and Debenedetti]{Torquato2000}
S.~Torquato, T.~M. Truskett and P.~G. Debenedetti, \emph{Phys. Rev. Lett.}, 2000, \textbf{84}, 2064\relax
\mciteBstWouldAddEndPuncttrue
\mciteSetBstMidEndSepPunct{\mcitedefaultmidpunct}
{\mcitedefaultendpunct}{\mcitedefaultseppunct}\relax
\EndOfBibitem
\bibitem[O'Hern \emph{et~al.}(2003)O'Hern, Silbert, Liu, and Nagel]{OHern2003}
C.~S. O'Hern, L.~E. Silbert, A.~J. Liu and S.~R. Nagel, \emph{Physical Review E}, 2003, \textbf{68}, 1--19\relax
\mciteBstWouldAddEndPuncttrue
\mciteSetBstMidEndSepPunct{\mcitedefaultmidpunct}
{\mcitedefaultendpunct}{\mcitedefaultseppunct}\relax
\EndOfBibitem
\bibitem[Luding(2016)]{Luding2016}
S.~Luding, \emph{Nature Physics}, 2016, \textbf{12}, 531--532\relax
\mciteBstWouldAddEndPuncttrue
\mciteSetBstMidEndSepPunct{\mcitedefaultmidpunct}
{\mcitedefaultendpunct}{\mcitedefaultseppunct}\relax
\EndOfBibitem
\bibitem[Chaudhuri \emph{et~al.}(2010)Chaudhuri, Berthier, and Sastry]{Chaudhuri2010}
P.~Chaudhuri, L.~Berthier and S.~Sastry, \emph{Phys. Rev. Lett.}, 2010, \textbf{104}, 165701\relax
\mciteBstWouldAddEndPuncttrue
\mciteSetBstMidEndSepPunct{\mcitedefaultmidpunct}
{\mcitedefaultendpunct}{\mcitedefaultseppunct}\relax
\EndOfBibitem
\bibitem[Bertrand \emph{et~al.}(2016)Bertrand, Behringer, Chakraborty, O'Hern, and Shattuck]{Bertrand2016}
T.~Bertrand, R.~P. Behringer, B.~Chakraborty, C.~S. O'Hern and M.~D. Shattuck, \emph{Physical Review E}, 2016, \textbf{93}, 1--7\relax
\mciteBstWouldAddEndPuncttrue
\mciteSetBstMidEndSepPunct{\mcitedefaultmidpunct}
{\mcitedefaultendpunct}{\mcitedefaultseppunct}\relax
\EndOfBibitem
\bibitem[Onoda and Liniger(1990)]{Onoda1990}
G.~Y. Onoda and E.~G. Liniger, \emph{Physical Review Letters}, 1990, \textbf{64}, 2727--2730\relax
\mciteBstWouldAddEndPuncttrue
\mciteSetBstMidEndSepPunct{\mcitedefaultmidpunct}
{\mcitedefaultendpunct}{\mcitedefaultseppunct}\relax
\EndOfBibitem
\bibitem[Silbert(2010)]{Silbert2010}
L.~E. Silbert, \emph{Soft Matter}, 2010, \textbf{6}, 2918--2924\relax
\mciteBstWouldAddEndPuncttrue
\mciteSetBstMidEndSepPunct{\mcitedefaultmidpunct}
{\mcitedefaultendpunct}{\mcitedefaultseppunct}\relax
\EndOfBibitem
\bibitem[Santos \emph{et~al.}(2020)Santos, Bolintineanu, Grest, Lechman, Plimpton, Srivastava, and Silbert]{santosGranularPackingsSliding2020}
A.~P. Santos, D.~S. Bolintineanu, G.~S. Grest, J.~B. Lechman, S.~J. Plimpton, I.~Srivastava and L.~E. Silbert, \emph{Phys. Rev. E}, 2020, \textbf{102}, 032903\relax
\mciteBstWouldAddEndPuncttrue
\mciteSetBstMidEndSepPunct{\mcitedefaultmidpunct}
{\mcitedefaultendpunct}{\mcitedefaultseppunct}\relax
\EndOfBibitem
\bibitem[Song \emph{et~al.}(2008)Song, Wang, and Makse]{Song2008a}
C.~Song, P.~Wang and H.~A. Makse, \emph{Nature}, 2008, \textbf{453}, 629--632\relax
\mciteBstWouldAddEndPuncttrue
\mciteSetBstMidEndSepPunct{\mcitedefaultmidpunct}
{\mcitedefaultendpunct}{\mcitedefaultseppunct}\relax
\EndOfBibitem
\bibitem[Silbert \emph{et~al.}(2002)Silbert, Ertaş, Grest, Halsey, and Levine]{Silbert2002}
L.~E. Silbert, D.~Ertaş, G.~S. Grest, T.~C. Halsey and D.~Levine, \emph{Physical Review E}, 2002, \textbf{65}, 1--6\relax
\mciteBstWouldAddEndPuncttrue
\mciteSetBstMidEndSepPunct{\mcitedefaultmidpunct}
{\mcitedefaultendpunct}{\mcitedefaultseppunct}\relax
\EndOfBibitem
\bibitem[Shundyak \emph{et~al.}(2007)Shundyak, {Van Hecke}, and {Van Saarloos}]{Shundyak2007}
K.~Shundyak, M.~{Van Hecke} and W.~{Van Saarloos}, \emph{Physical Review E}, 2007, \textbf{75}, 010301\relax
\mciteBstWouldAddEndPuncttrue
\mciteSetBstMidEndSepPunct{\mcitedefaultmidpunct}
{\mcitedefaultendpunct}{\mcitedefaultseppunct}\relax
\EndOfBibitem
\bibitem[Somfai \emph{et~al.}(2007)Somfai, {Van Hecke}, Ellenbroek, Shundyak, and {Van Saarloos}]{Somfai2007}
E.~Somfai, M.~{Van Hecke}, W.~G. Ellenbroek, K.~Shundyak and W.~{Van Saarloos}, \emph{Physical Review E}, 2007, \textbf{75}, 020301\relax
\mciteBstWouldAddEndPuncttrue
\mciteSetBstMidEndSepPunct{\mcitedefaultmidpunct}
{\mcitedefaultendpunct}{\mcitedefaultseppunct}\relax
\EndOfBibitem
\bibitem[Bi \emph{et~al.}(2011)Bi, Zhang, Chakraborty, and Behringer]{Bi2011}
D.~Bi, J.~Zhang, B.~Chakraborty and R.~P. Behringer, \emph{Nature}, 2011, \textbf{480}, 355--358\relax
\mciteBstWouldAddEndPuncttrue
\mciteSetBstMidEndSepPunct{\mcitedefaultmidpunct}
{\mcitedefaultendpunct}{\mcitedefaultseppunct}\relax
\EndOfBibitem
\bibitem[Farrell \emph{et~al.}(2010)Farrell, Martini, and Menon]{Farrell2010}
G.~R. Farrell, K.~M. Martini and N.~Menon, \emph{Soft Matter}, 2010, \textbf{6}, 2925--2930\relax
\mciteBstWouldAddEndPuncttrue
\mciteSetBstMidEndSepPunct{\mcitedefaultmidpunct}
{\mcitedefaultendpunct}{\mcitedefaultseppunct}\relax
\EndOfBibitem
\bibitem[Delaney \emph{et~al.}(2011)Delaney, Hilton, and Cleary]{Delaney2011}
G.~W. Delaney, J.~E. Hilton and P.~W. Cleary, \emph{Physical Review E}, 2011, \textbf{83}, 051305\relax
\mciteBstWouldAddEndPuncttrue
\mciteSetBstMidEndSepPunct{\mcitedefaultmidpunct}
{\mcitedefaultendpunct}{\mcitedefaultseppunct}\relax
\EndOfBibitem
\bibitem[Hoy and Kr{\"o}ger(2020)]{hoyUnifiedAnalyticExpressions2020}
R.~S. Hoy and M.~Kr{\"o}ger, \emph{Phys. Rev. Lett.}, 2020, \textbf{124}, 147801\relax
\mciteBstWouldAddEndPuncttrue
\mciteSetBstMidEndSepPunct{\mcitedefaultmidpunct}
{\mcitedefaultendpunct}{\mcitedefaultseppunct}\relax
\EndOfBibitem
\bibitem[Bililign \emph{et~al.}(2019)Bililign, Kollmer, and Daniels]{Bililign2019}
E.~S. Bililign, J.~E. Kollmer and K.~E. Daniels, \emph{Physical Review Letters}, 2019, \textbf{122}, 38001\relax
\mciteBstWouldAddEndPuncttrue
\mciteSetBstMidEndSepPunct{\mcitedefaultmidpunct}
{\mcitedefaultendpunct}{\mcitedefaultseppunct}\relax
\EndOfBibitem
\bibitem[Lubachevsky and Stillinger(1990)]{Lubachevsky1990}
B.~D. Lubachevsky and F.~H. Stillinger, \emph{Journal of Statistical Physics}, 1990, \textbf{60}, 561--583\relax
\mciteBstWouldAddEndPuncttrue
\mciteSetBstMidEndSepPunct{\mcitedefaultmidpunct}
{\mcitedefaultendpunct}{\mcitedefaultseppunct}\relax
\EndOfBibitem
\bibitem[O'Hern \emph{et~al.}(2002)O'Hern, Langer, Liu, and Nagel]{OHern2002}
C.~S. O'Hern, S.~A. Langer, A.~J. Liu and S.~R. Nagel, \emph{Physical Review Letters}, 2002, \textbf{88}, 075507\relax
\mciteBstWouldAddEndPuncttrue
\mciteSetBstMidEndSepPunct{\mcitedefaultmidpunct}
{\mcitedefaultendpunct}{\mcitedefaultseppunct}\relax
\EndOfBibitem
\bibitem[Charbonneau \emph{et~al.}(2012)Charbonneau, Corwin, Parisi, and Zamponi]{Charbonneau2012}
P.~Charbonneau, E.~I. Corwin, G.~Parisi and F.~Zamponi, \emph{Physical Review Letters}, 2012, \textbf{109}, 205501\relax
\mciteBstWouldAddEndPuncttrue
\mciteSetBstMidEndSepPunct{\mcitedefaultmidpunct}
{\mcitedefaultendpunct}{\mcitedefaultseppunct}\relax
\EndOfBibitem
\bibitem[Srivastava \emph{et~al.}(2019)Srivastava, Silbert, Grest, and Lechman]{Srivastava2019}
I.~Srivastava, L.~E. Silbert, G.~S. Grest and J.~B. Lechman, \emph{Physical Review Letters}, 2019, \textbf{122}, 48003\relax
\mciteBstWouldAddEndPuncttrue
\mciteSetBstMidEndSepPunct{\mcitedefaultmidpunct}
{\mcitedefaultendpunct}{\mcitedefaultseppunct}\relax
\EndOfBibitem
\bibitem[Clemmer \emph{et~al.}(2021)Clemmer, Srivastava, Grest, and Lechman]{clemmerShearNotAlways2021}
J.~T. Clemmer, I.~Srivastava, G.~S. Grest and J.~B. Lechman, \emph{Phys. Rev. Lett.}, 2021, \textbf{127}, 268003\relax
\mciteBstWouldAddEndPuncttrue
\mciteSetBstMidEndSepPunct{\mcitedefaultmidpunct}
{\mcitedefaultendpunct}{\mcitedefaultseppunct}\relax
\EndOfBibitem
\bibitem[Kohlrausch(1854)]{Kohlrausch1854}
R.~Kohlrausch, \emph{Pogg. Ann. Phys. Chem.}, 1854, \textbf{91}, 179--214\relax
\mciteBstWouldAddEndPuncttrue
\mciteSetBstMidEndSepPunct{\mcitedefaultmidpunct}
{\mcitedefaultendpunct}{\mcitedefaultseppunct}\relax
\EndOfBibitem
\bibitem[Williams and Watts(1970)]{Williams1970}
G.~Williams and D.~C. Watts, \emph{Trans. Faraday Soc.}, 1970, \textbf{66}, 80--85\relax
\mciteBstWouldAddEndPuncttrue
\mciteSetBstMidEndSepPunct{\mcitedefaultmidpunct}
{\mcitedefaultendpunct}{\mcitedefaultseppunct}\relax
\EndOfBibitem
\bibitem[Knight \emph{et~al.}(1995)Knight, Fandrich, {Ning Lau}, Jaeger, and Nagel]{Knight1995}
J.~B. Knight, C.~G. Fandrich, C.~{Ning Lau}, H.~M. Jaeger and S.~R. Nagel, \emph{Physical Review E}, 1995, \textbf{51}, 3957--3963\relax
\mciteBstWouldAddEndPuncttrue
\mciteSetBstMidEndSepPunct{\mcitedefaultmidpunct}
{\mcitedefaultendpunct}{\mcitedefaultseppunct}\relax
\EndOfBibitem
\bibitem[Philippe and Bideau(2002)]{Philippe2002}
P.~Philippe and D.~Bideau, \emph{Euro. Phys. Lett.}, 2002, \textbf{60}, 677--683\relax
\mciteBstWouldAddEndPuncttrue
\mciteSetBstMidEndSepPunct{\mcitedefaultmidpunct}
{\mcitedefaultendpunct}{\mcitedefaultseppunct}\relax
\EndOfBibitem
\bibitem[Richard \emph{et~al.}(2005)Richard, Nicodemi, Delannay, Ribi{\`{e}}re, and Bideau]{Richard2005}
P.~Richard, M.~Nicodemi, R.~Delannay, P.~Ribi{\`{e}}re and D.~Bideau, \emph{Nature materials}, 2005, \textbf{4}, 121--128\relax
\mciteBstWouldAddEndPuncttrue
\mciteSetBstMidEndSepPunct{\mcitedefaultmidpunct}
{\mcitedefaultendpunct}{\mcitedefaultseppunct}\relax
\EndOfBibitem
\bibitem[Rosato \emph{et~al.}(2010)Rosato, Dybenko, Horntrop, Ratnaswamy, Kondic, and Carlo]{Rosato2010}
A.~D. Rosato, O.~Dybenko, D.~J. Horntrop, V.~Ratnaswamy, L.~Kondic and M.~Carlo, \emph{Phys. Rev. E}, 2010, \textbf{81}, 061301\relax
\mciteBstWouldAddEndPuncttrue
\mciteSetBstMidEndSepPunct{\mcitedefaultmidpunct}
{\mcitedefaultendpunct}{\mcitedefaultseppunct}\relax
\EndOfBibitem
\bibitem[Kumar and Luding(2016)]{Kumar2016}
N.~Kumar and S.~Luding, \emph{Granular Matter}, 2016, \textbf{18}, 58\relax
\mciteBstWouldAddEndPuncttrue
\mciteSetBstMidEndSepPunct{\mcitedefaultmidpunct}
{\mcitedefaultendpunct}{\mcitedefaultseppunct}\relax
\EndOfBibitem
\bibitem[Dagois-Bohy \emph{et~al.}(2012)Dagois-Bohy, Tighe, Simon, Henkes, and {Van Hecke}]{Dagois-Bohy2012}
S.~Dagois-Bohy, B.~P. Tighe, J.~Simon, S.~Henkes and M.~{Van Hecke}, \emph{Physical Review Letters}, 2012, \textbf{109}, 1--5\relax
\mciteBstWouldAddEndPuncttrue
\mciteSetBstMidEndSepPunct{\mcitedefaultmidpunct}
{\mcitedefaultendpunct}{\mcitedefaultseppunct}\relax
\EndOfBibitem
\bibitem[Smith \emph{et~al.}(2014)Smith, Srivastava, Fisher, and Alam]{Smith2014}
K.~C. Smith, I.~Srivastava, T.~S. Fisher and M.~Alam, \emph{Physical Review E}, 2014, \textbf{89}, 042203\relax
\mciteBstWouldAddEndPuncttrue
\mciteSetBstMidEndSepPunct{\mcitedefaultmidpunct}
{\mcitedefaultendpunct}{\mcitedefaultseppunct}\relax
\EndOfBibitem
\bibitem[Thompson \emph{et~al.}(2022)Thompson, Aktulga, Berger, Bolintineanu, Brown, Crozier, {in 't Veld}, Kohlmeyer, Moore, Nguyen, Shan, Stevens, Tranchida, Trott, and Plimpton]{thompsonLAMMPSFlexibleSimulation2022}
A.~P. Thompson, H.~M. Aktulga, R.~Berger, D.~S. Bolintineanu, W.~M. Brown, P.~S. Crozier, P.~J. {in 't Veld}, A.~Kohlmeyer, S.~G. Moore, T.~D. Nguyen, R.~Shan, M.~J. Stevens, J.~Tranchida, C.~Trott and S.~J. Plimpton, \emph{Computer Physics Communications}, 2022, \textbf{271}, 108171\relax
\mciteBstWouldAddEndPuncttrue
\mciteSetBstMidEndSepPunct{\mcitedefaultmidpunct}
{\mcitedefaultendpunct}{\mcitedefaultseppunct}\relax
\EndOfBibitem
\bibitem[Shinoda \emph{et~al.}(2004)Shinoda, Shiga, and Mikami]{Shinoda2004}
W.~Shinoda, M.~Shiga and M.~Mikami, \emph{Physical Review B}, 2004, \textbf{69}, 16--18\relax
\mciteBstWouldAddEndPuncttrue
\mciteSetBstMidEndSepPunct{\mcitedefaultmidpunct}
{\mcitedefaultendpunct}{\mcitedefaultseppunct}\relax
\EndOfBibitem
\bibitem[Parrinello and Rahman(1981)]{Parrinello1981}
M.~Parrinello and A.~Rahman, \emph{J. Appl. Phys.}, 1981, \textbf{52}, 7182\relax
\mciteBstWouldAddEndPuncttrue
\mciteSetBstMidEndSepPunct{\mcitedefaultmidpunct}
{\mcitedefaultendpunct}{\mcitedefaultseppunct}\relax
\EndOfBibitem
\bibitem[Martyna \emph{et~al.}(1994)Martyna, Tobias, and Klein]{Martyna1994}
G.~J. Martyna, D.~J. Tobias and M.~L. Klein, \emph{J. Chem. Phys.}, 1994, \textbf{101}, 4177--4189\relax
\mciteBstWouldAddEndPuncttrue
\mciteSetBstMidEndSepPunct{\mcitedefaultmidpunct}
{\mcitedefaultendpunct}{\mcitedefaultseppunct}\relax
\EndOfBibitem
\bibitem[Reddy \emph{et~al.}(1992)Reddy, Saxena, and Budiman]{reddyDevelopmentTrueTriaxial1992}
K.~R. Reddy, S.~K. Saxena and J.~S. Budiman, \emph{Geotech. Test. J.}, 1992, \textbf{15}, 89--105\relax
\mciteBstWouldAddEndPuncttrue
\mciteSetBstMidEndSepPunct{\mcitedefaultmidpunct}
{\mcitedefaultendpunct}{\mcitedefaultseppunct}\relax
\EndOfBibitem
\bibitem[Nan and Hoy(2023)]{nan_ultraslow_2023}
K.~Nan and R.~S. Hoy, \emph{Phys. Rev. Lett.}, 2023, \textbf{130}, 166102\relax
\mciteBstWouldAddEndPuncttrue
\mciteSetBstMidEndSepPunct{\mcitedefaultmidpunct}
{\mcitedefaultendpunct}{\mcitedefaultseppunct}\relax
\EndOfBibitem
\end{mcitethebibliography}
\bibliographystyle{rsc} 

\section{Supplementary Material: Initial kinetic energy impact on approach to packing}\label{sec:SIkeapproach}
\begin{figure}[!htbp]
\centering	
\includegraphics[width=12cm]{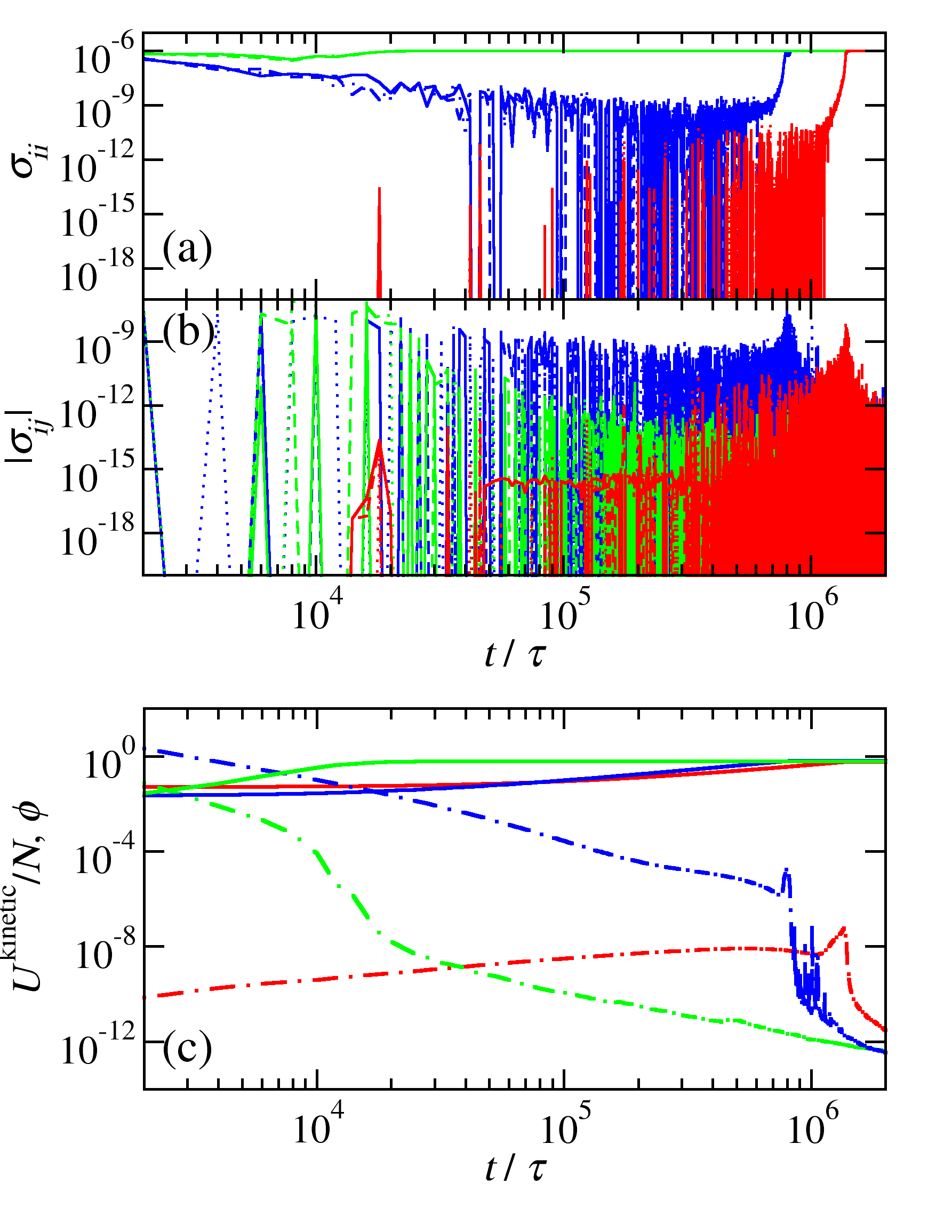}
\caption{The (a) diagonal and (b) off-diagonal components of the applied stress tensor for $P_{a}=10^{-6}$ for three different initial pressures $P_0=0$ (red) $P_0=10^{-4}$ (blue) and $P_0=10^{-2}$ (green). The different components of the stress tensor are plotted as different line types: xx, xy (solid lines), yy, xz (dashed lines) and zz, yz (dotted lines). The off-diagonal components of the stress tensor are shown as averages over 10 timesteps for clarity.
(c) The kinetic energy (dot-dashed lines) and the volume fraction (solid) as a function of time For (a), (b) and (c) the box parameters are $P_{\text{damp}}=2.25$ and $f_{\text{drag}}=0$, and the friction state is $\mu_s=0.2$.}
\label{fig:SIstresses}
\end{figure}

Figure~\ref{fig:SIstresses} shows the role of the initial pressure on the approach to jamming in stress and microstructural properties. The figure in the main article, Figure 2, has box drag $f_{\text{drag}}=0.2$, while Figure~\ref{fig:SIstresses} has no box drag $f_{\text{drag}}=0$ and different pressure. The time to pack increases and fluctuations decrease with increasing $f_{\text{drag}}$ and decreasing pressure. 
These large fluctuations make the system less numerically stable as $P_a$ decreases, which is why flow simulations, $\sigma_xy\ne 0$ for example, may need $f_{\text{drag}}\ne 0$.

\section{Supplementary Material: Non-monotonic volume fraction dependence for other methods}\label{sec:SImethodII}
\begin{figure}[!htbp]
\centering	
\includegraphics[width=10cm]{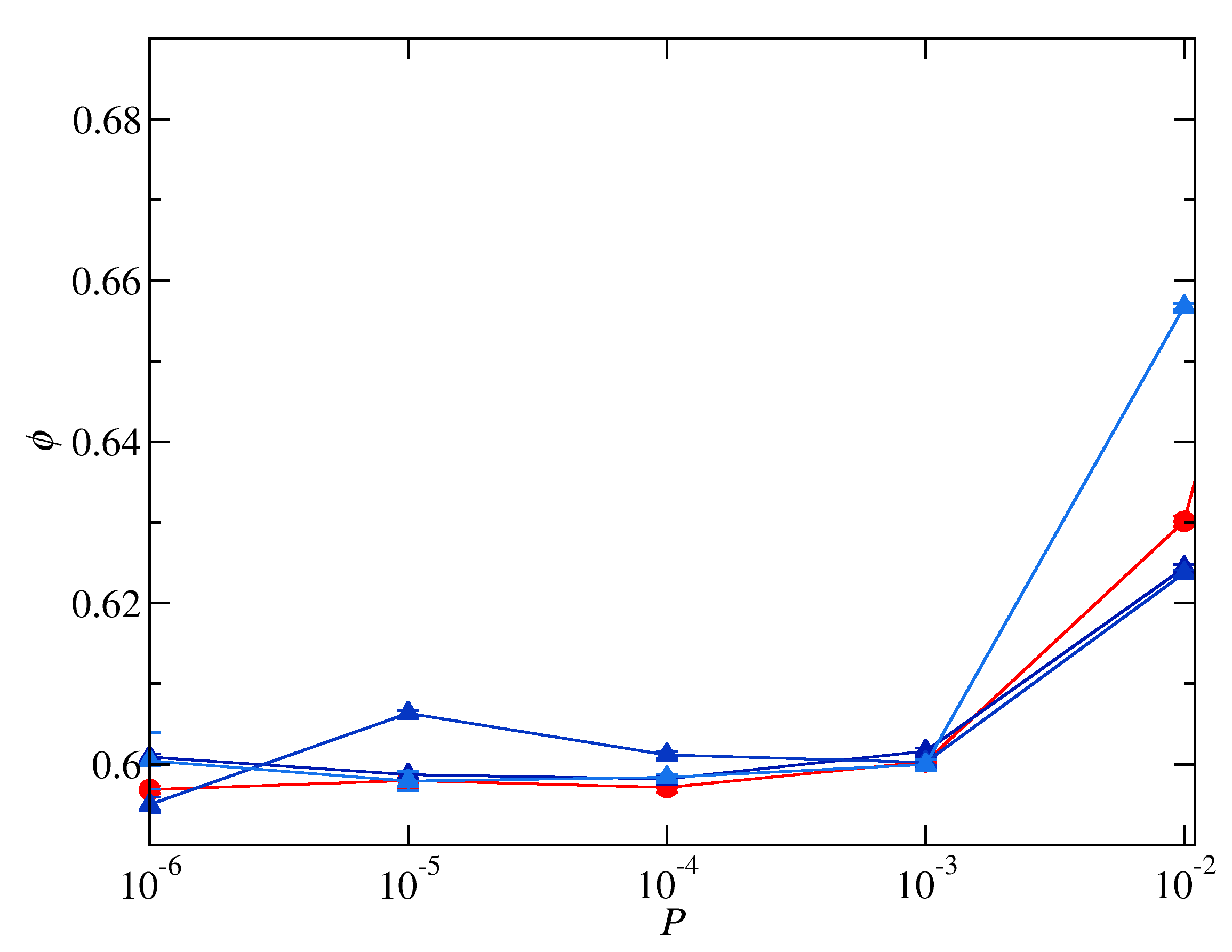}
\caption{Packing fraction as a function of the pressure $\phi(P_a)$ for method I (red circles) and II (blue triangles) with no initial pressure $P_0=0$. Method II, where $P_{a,0}$ is varied from $10^{-4}$ (dark blue triangles), $10^{-3}$ (medium blue triangles) and $10^{-1}$ (light blue triangles). The box parameters are $P_{\text{damp}}=2.25$ and $f_{\text{drag}}=0$. Uncertainties are calculated from 6 different packings of $N=10^4$ particles are similar in size to the symbols.}
\label{fig:SImethodsnoKE}
\end{figure}

\begin{figure}[!htbp]
\centering	
\includegraphics[width=10cm]{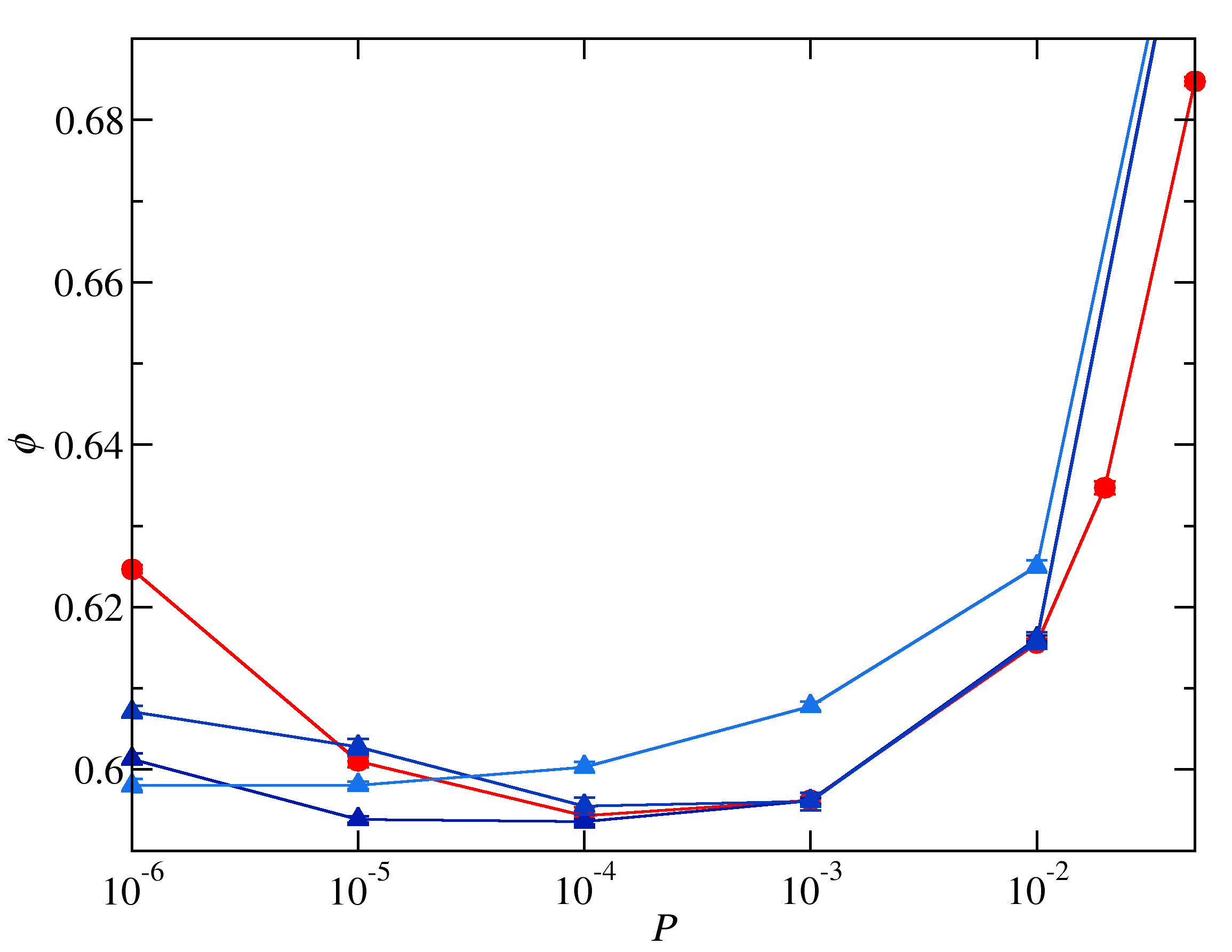}
\caption{Packing fraction as a function of the pressure $\phi(P_a)$ for method I (red circles) and II (blue triangles) with no initial pressure $P_0=10^{-2}$. Method II, where $P_{a,i}$ is varied from $10^{-4}$ (dark blue triangles), $10^{-3}$ (medium blue triangles) and $10^{-1}$ (light blue triangles). The box parameters are $P_{\text{damp}}=2.25$ and $f_{\text{drag}}=0$. Uncertainties are calculated from 6 different packings of $N=10^4$ particles are similar in size to the symbols.}
\label{fig:SImethodsKE}
\end{figure}

Figures~\ref{fig:SImethodsnoKE} and ~\ref{fig:SImethodsKE} show the role of the initial pressure on packing methods I and II. At $t=0$ packing method I starts at $\phi_0 = 0.05$, and a constant pressure $P_{a,f}$ is applied until the system jams. Packing method II first packs at an initial, high pressure $P_{a,0} > P_{a,f}$, and then the target pressure is instantaneously decreased to $P_{a,f}$. Figures~\ref{fig:SImethodsnoKE} and ~\ref{fig:SImethodsKE} also show packings with more overcompression pressures $P_{a,i}$ for comparison. These figures demonstrate the distance from jamming for both the volume fraction and pressure affect the final microstructure. 
\section{Supplementary Material:  Role of friction}\label{sec:SIfriction}
\begin{figure}[!htbp]
\centering
\includegraphics[width=10cm]{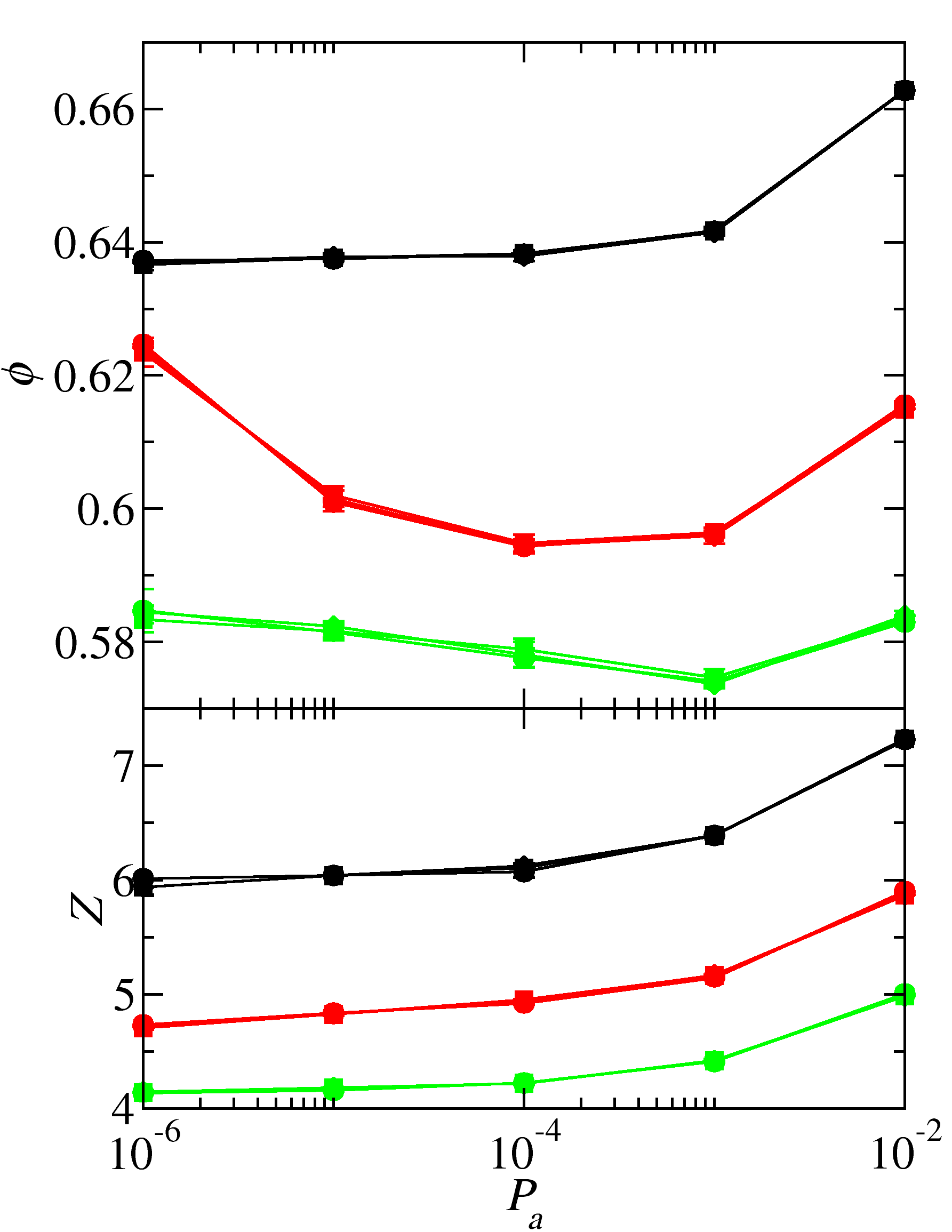}
\caption{Packing fraction $\phi$ (top) and average coordination number without rattlers $Z$ (bottom) as a function of the pressure for sliding frictions $\mu_s$ = 0.0 (black), 0.2 (red) and 1 (green) with non-zero initial pressure $P_0=1.5\times 10^{-2}$. Different constraints on the applied stress satisfied at jamming are shown: (i) $\sigma_{a,xx}=\sigma_{a,yy}=\sigma_{a,zz}=P$ and $\sigma_{a,xy}=\sigma_{a,yz}=\sigma_{a,xz}=0$ (circles), (ii) $(\sigma_{a,xx}+\sigma_{a,yy}+\sigma_{a,zz})/3=P$ and $\sigma_{a,xy}=\sigma_{a,yz}=\sigma_{a,xz}=0$ (squares), and (iii) $\sigma_{a,xx}=\sigma_{a,yy}=\sigma_{a,zz}=P$ in an perfect orthorombic cube, where $\sigma_{a,xy},\sigma_{a,yz}$ and $\sigma_{a,xz}$ are not set (diamonds). The symbols for the three jamming states overlap. These packings were generated with method I, $P_{\text{damp}}=2$ and $f_{\text{drag}}=1$. Uncertainties are calculated from 6 different packings of $N=10^4$ particles are similar in size to the symbols.}
\label{fig:SIPhiZPress}
\end{figure}

Figure~\ref{fig:SIPhiZPress} shows the role of friction on the $\phi(P_a)$ depth, where as the coordination number is monotonic regardless of friction and the initial pressure. Figure~\ref{fig:SIPhiZPress} also demonstrates that the exact stress-tensor definition does not  statistically change $\phi$ or $Z$.

\end{document}